\documentclass[11pt,a4paper]{article}
\usepackage[a4paper, margin=1in]{geometry}

\usepackage{amsmath, amssymb}
\usepackage{graphicx}
\usepackage{hyperref}
\usepackage{listings} 
\usepackage{natbib}
\usepackage{authblk}
\usepackage{comment}
\usepackage{algorithm}
\usepackage{algorithmic}
\usepackage{graphicx}
\usepackage{caption}
\usepackage{subcaption} 
\usepackage{xcolor}
\usepackage{float}

\title{FSIM: A Pedagogical and Extensible HPC Framework \\ for the Hartree–Fock Method}

\author[1]{Mario Hern{\'a}ndez Vera\thanks{ \href{mailto:mario.hernandezvera@lrz.de}{mario.hernandezvera@lrz.de} ~ \href{mailto:marhvera@gmail.com}{marhvera@gmail.com} }}
\affil[1]{Leibniz Supercomputing Centre, Germany}
\date{\today}

\begin{document}

\maketitle

\begin{abstract}
Efficient computation of molecular integrals and Hartree–Fock energy remains a central topic in 
quantum-chemistry algorithm development. Although many 
sophisticated open-source packages are available, understanding their implementations 
from first principles can be difficult for students and developers alike.
In this work, we present a concise overview and an extensible pedagogical framework that implement 
the Hartree–Fock method and the McMurchie-Davidson scheme for molecular integral evaluation. 
The implementation follows an 
object-oriented design in C++, emphasizing clarity and modularity. 
We also discuss 
strategies for parallel execution, including distributed computing with MPI and 
shared-memory parallelization with OpenMP. 
Beyond presenting a working reference, this work establishes a learning platform for 
further exploration, including suggested mini-projects for algorithmic optimization 
and HPC scalability.
The accompanying open-source library, FSIM, described in this work, serves as a compact resource 
for teaching and research in computational chemistry and high-performance computing.
\end{abstract}

\section{Introduction}

The Hartree–Fock (HF) method remains one of the cornerstones of quantum chemistry \cite{Helgaker2000}. 
Together with the Born–Oppenheimer (BO) approximation, it provides a powerful and widely used framework for describing the electronic structure of molecules. 
Although the HF model neglects static and dynamic electron correlation, it yields molecular orbitals that serve as the reference basis for more accurate post-Hartree–Fock approaches. 
For this reason, reliable and efficient HF implementations are fundamental components of nearly all modern electronic-structure codes.

A major source of efficiency in contemporary HF programs arises from the use of Gaussian-type orbitals (GTOs) as basis functions. 
The analytic properties of Gaussians enable systematic and efficient evaluation of molecular integrals, allowing calculations to scale to systems containing hundreds or even thousands of atoms. 
With appropriate integral-screening and parallelization techniques, the HF method has become a key workload on modern high-performance computing (HPC) systems. 
Indeed, many demonstrations of large-scale HPC performance have used quantum-chemistry benchmarks—often based on HF or related methods—to showcase parallel scalability and computational efficiency \cite{stocks2024hfexascale}.

Beyond its scientific importance, the HF method continues to hold significant \emph{educational and technological value}. 
As computer architectures evolve—through multicore processors, GPUs, and other accelerators—quantum-chemistry algorithms must be adapted to exploit these resources effectively \cite{Gordon2020}. 
Open, modular implementations allow students and developers to experiment with data structures, memory layouts, and parallel strategies, fostering a deeper understanding of algorithmic bottlenecks and opportunities for optimization. 
The transparent formulation of HF theory and its mapping to modern HPC paradigms therefore provide a fertile training ground at the intersection of \emph{chemistry, physics, and computer science}.

Despite this potential, relatively few open-source, pedagogically oriented C++ implementations combine 
high-performance computing features with a clear exposition of the theoretical and computational structure 
of the HF method. Most mature quantum-chemistry packages are large, complex, and heavily optimized for 
performance rather than clarity, making them difficult to study and extend. 
To address this gap, the first part of this paper provides a concise pedagogical review of the 
HF formalism and the McMurchie–Davidson (McD) scheme for Gaussian integral evaluation~\cite{MCMURCHIE1978218}, 
establishing the theoretical foundation for implementation.

Building on this foundation, we introduce FSIM—a minimal, open-source C++ library that implements the 
restricted HF method using the McD integral scheme and 
supports parallelization through MPI and OpenMP, designed primarily as a pedagogical demonstration rather 
than a production-scale code. FSIM is designed as a modular and extensible framework that bridges theory and implementation, 
serving as both a pedagogical tool and a platform for further exploration. 
The accompanying mini-projects and open design aim to encourage students to extend the code, 
explore optimization strategies, and contribute to future high-performance developments.
We would like to emphasize that this work is not positioned as a production electronic-structure 
package, but rather as a minimal reference implementation intended for educational 
use and software review.

The remainder of this paper is organized to bridge theoretical foundations with practical implementation 
and learning opportunities. Section~\ref{sec:math_theory} introduces the mathematical formulation of the HF 
method, beginning with the definition of Gaussian basis functions and progressing through the evaluation of 
molecular integrals. This theoretical groundwork provides the context for Section~\ref{sec:implementation}, which describes the 
design and structure of the FSIM library.
Here, we outline the software architecture, the self-consistent field (SCF) solver, and the integral 
engine implementation.
Section~\ref{sec:validation} presents validation tests and performance results, while Section~\ref{sec:profiling} 
analyzes the profiling of both serial and parallel components of the code.
To highlight its educational intent, Section~\ref{sec:projects} proposes a series of guided mini-projects 
that extend the framework toward more advanced optimization and scalability studies.

\section{Mathematical and Theoretical Formulation }
\label{sec:math_theory}

The mathematical formalism of the Hartree–Fock method and the McMurchie–Davidson 
scheme for integrals have been described in detail by many authors, for example in the 
well-known textbook Ref.~\cite{Helgaker2000}. Here, we include a summary of key concepts 
and equations for consistency within this paper.

\subsection{Basis Functions}

\subsubsection{Contracted Gaussian-type orbitals}

Molecular orbitals (MOs) are commonly expressed as linear combinations of atomic orbitals (AOs), 
following the linear combination of atomic orbitals (LCAO) framework. 
This representation forms the foundation 
of most electronic-structure methods, as it allows the many-electron Schrödinger 
equation to be reformulated in terms of a finite basis set:
\begin{equation}
\psi_{p}(\mathbf{r}) = \sum_{\mu} C_{\mu p}\, \phi_{\mu}(\mathbf{r}),
\label{eq:lcao}
\end{equation}
where $\phi_{\mu}(\mathbf{r})$ are the atomic basis functions and $C_{\mu p}$ are the molecular orbital coefficients.

In quantum chemistry calculations, atomic basis functions are usually represented by
\emph{contracted Gaussian-type orbitals} (CGTOs) with real solid harmonic angular dependence.
A CGTO centered on the nucleus at $\mathbf{A}$ is defined as a fixed linear combination
of primitive real solid harmonic Gaussian functions:
\begin{equation}
\phi_{\mu}^{l m}(\mathbf{r}) = \sum_{k} d_{k}^{\mu}\,
G_{l m}(\mathbf{r}, \alpha_{k}, \mathbf{A}),
\end{equation}
where $(l,m)$ are angular-momentum quantum numbers, $d_{k}^{\mu}$ are contraction coefficients,
and $\alpha_{k}$ are primitive Gaussian exponents.

Each primitive real solid harmonic Gaussian function is given by
\begin{equation}
G_{l m}(\mathbf{r}, \alpha, \mathbf{A}) =
S_{l m}(\mathbf{r}-\mathbf{A})\, e^{-\alpha|\mathbf{r}-\mathbf{A}|^{2}},
\end{equation}
where $S_{l m}(\mathbf{r}-\mathbf{A}) = r_{A}^{l}\, Y_{l m}(\widehat{\mathbf{r}-\mathbf{A}})$
is a real solid harmonic constructed from the spherical harmonics $Y_{l m}$.
For normalization, each primitive function is multiplied by a factor $N_{l m}(\alpha)$
such that $\langle G_{l m} | G_{l m} \rangle = 1$.

For the purpose of integral evaluation, it is convenient to expand
each primitive real solid harmonic Gaussian in terms of primitive
\emph{Cartesian Gaussian-type orbitals} \cite{Helgaker2000}:
\begin{equation}
G_{l m}(\mathbf{r}, \alpha, \mathbf{A})
= \sum_{|i|=l} S_{i}^{l m}\,
(x-A_{x})^{i_{x}}(y-A_{y})^{i_{y}}(z-A_{z})^{i_{z}}
e^{-\alpha|\mathbf{r}-\mathbf{A}|^{2}},
\end{equation}
where $i = (i_x, i_y, i_z)$ is a multi-index with $|i| = i_x+i_y+i_z = l$,
and $S_{i}^{l m}$ are the solid harmonic to Cartesian transformation coefficients, whose explicit 
expressions are given in Ref.~\cite{Helgaker2000}.

The final expansion of a CGTO in the Cartesian primitive basis can be written as
\begin{equation}
\phi_{\mu}^{lm}(\mathbf{r}) =
\sum_{k}^{K} \sum_{\substack{i_x, i_y, i_z}}
d_{k}^{\mu}\, S_{i_x, i_y, i_z}^{l m }\,
(x-A_{x})^{i_x}(y-A_{y})^{i_y}(z-A_{z})^{i_z}
e^{-\alpha_{k}^{\mu}|\mathbf{r}-\mathbf{A}_{\mu}|^{2}}.
\end{equation}

The expansion in  Cartesian primitive basis Gaussians
increases the number of integrals to be evaluated,
but enables highly efficient computation because all Gaussian integrals
are analytically tractable and can be evaluated through recurrence relations \cite{Helgaker2000, 2012Reine}.

\subsubsection{Cartesian Gaussian Functions}

It is useful to briefly review the definition of a primitive Cartesian Gaussian-type orbital, as these functions form a central component of the basis expansion and molecular integral evaluation.
One such function centered at $\mathbf{A}$ is defined as
\begin{equation}
G_{i_x i_y i_z}(\mathbf{r};\alpha,\mathbf{A}) =
(x - A_x)^{i_x}(y - A_y)^{i_y}(z - A_z)^{i_z}
e^{-\alpha |\mathbf{r}-\mathbf{A}|^2},
\end{equation}
where $\alpha>0$ is the orbital exponent, $(i_x,i_y,i_z)$ are nonnegative
integers, and the total angular-momentum quantum number is
$\ell = i_x + i_y + i_z$.

A Cartesian Gaussian is separable in the Cartesian directions:
\begin{equation}
G_{i_x i_y i_z}(\alpha,\mathbf{r}_A)
= G_{i_x}(\alpha,x_A)\, G_{i_y}(\alpha,y_A)\, G_{i_z}(\alpha,z_A),
\end{equation}
where, for example, the $x$ component is
\begin{equation}
G_{i_x}(\alpha,x_A) = x_A^{i_x}\, e^{-\alpha x_A^2},
\quad \text{with } x_A = x - A_x.
\end{equation}

The main advantage of Gaussians is their simple algebraic properties.  
The \emph{Gaussian product theorem} states that the product of two
spherical Gaussians centered at $\mathbf{A}$ and $\mathbf{B}$ with exponents
$\alpha$ and $\beta$ is itself a spherical Gaussian centered at an intermediate
point $\mathbf{P}$:
\begin{equation}
\mathbf{P} = \frac{\alpha \mathbf{A} + \beta \mathbf{B}}{\alpha + \beta},
\quad
p = \alpha + \beta.
\end{equation}
The product can then be written as
\begin{equation}
e^{-\alpha|\mathbf{r}-\mathbf{A}|^2}
e^{-\beta|\mathbf{r}-\mathbf{B}|^2}
= \kappa_{AB}\, e^{-p|\mathbf{r}-\mathbf{P}|^2},
\end{equation}
where $\kappa_{AB} = \exp(-\mu_{AB}|\mathbf{A}-\mathbf{B}|^2)$
and $\mu_{AB} = \frac{\alpha\beta}{\alpha+\beta}$.
This property allows the reduction of two-center integrals to one-center integrals (as in the case of overlap integrals), thereby simplifying their numerical evaluation within quantum-chemical algorithms~\cite{Helgaker2000}.

\subsubsection{Hermite Gaussians}

As will be shown in the following sections, expansion over Cartesian Gaussian 
functions alone is insufficient for the efficient evaluation of one- and two-electron integrals.
Within the McD scheme, it is therefore convenient to introduce the \emph{Hermite Gaussian functions}, 
which are obtained by taking derivatives of a spherical Gaussian with respect to its center~$\mathbf{P}$:
\begin{equation}
\Lambda_{tuv}(\mathbf{r};p,\mathbf{P}) =
\left(\frac{\partial}{\partial P_x}\right)^{t}
\left(\frac{\partial}{\partial P_y}\right)^{u}
\left(\frac{\partial}{\partial P_z}\right)^{v}
e^{-p|\mathbf{r}-\mathbf{P}|^2},
\end{equation}
where $t,u,v \ge 0$ are integers and $p>0$ is the Gaussian exponent.

Like Cartesian Gaussians, Hermite Gaussians factorize along Cartesian
directions:
\begin{equation}
\Lambda_{tuv}(\mathbf{r};p,\mathbf{P})
= \Lambda_t(x_P)\, \Lambda_u(y_P)\, \Lambda_v(z_P),
\end{equation}
with, for instance, 
\begin{equation}
\Lambda_t(x_P) =
\left(\frac{\partial}{\partial P_x}\right)^t e^{-p x_P^2},
\qquad x_P = x - P_x.
\end{equation}
This separability simplifies the evaluation of integrals and the development
of recurrence relations. Hermite Gaussians satisfy simple recurrence and differentiation relations.
In one dimension,
\begin{equation}
x_P \Lambda_t(x_P)
= \frac{1}{2p}\Lambda_{t+1}(x_P) + t\,\Lambda_{t-1}(x_P),
\label{eq:hermite-recurrence}
\end{equation}
which allows polynomial prefactors to be generated recursively.

They also possess an especially simple integration property:
\begin{equation}
\int_{-\infty}^{\infty} \Lambda_t(x_P)\,dx
= \delta_{t0}\sqrt{\frac{\pi}{p}},
\end{equation}
so that only Hermite $s$-functions ($t=u=v=0$) contribute to
one-dimensional overlap integrals.

The fundamental application of Hermite Gaussians in the McD scheme is to expand products of 
Cartesian Gaussians, also known as overlap distributions. For instance, along the $x$-axis,
\begin{equation}
(x - A_x)^i (x - B_x)^j e^{-p(x - P_x)^2}
= \sum_{t=0}^{i+j} E_{t}^{ij}\, \Lambda_t(x_P),
\end{equation}
where the expansion coefficients $E_{t}^{ij}$ can be evaluated recursively \cite{Helgaker2000}:
\begin{align}
E_{t}^{i+1,j} &=
\frac{1}{2p} E_{\,t-1}^{ij}
+ (P_x - A_x)\, E_{ij}^{\,t}
+ (t+1)\, E_{\,t+1}^{ij}, \label{eq:recurrence-A}\\[4pt]
E_{t}^{i,j+1} &=
\frac{1}{2p} E_{\,t-1}^{ij}
+ (P_x - B_x)\, E_{t}^{ij}
+ (t+1)\, E_{\,t+1}^{ij}, \label{eq:recurrence-B}
\end{align}
with the initial and boundary conditions
\begin{align}
E_{0}^{00} &= \exp\!\left[-\frac{\alpha\beta}{\alpha+\beta}\,(\!A_x - B_x\!)^2\right], \\[4pt]
E_{t}^{ij} &= 0 \quad \text{if } t < 0 \text{ or } t > i+j.
\end{align}
Analogous relations hold for the $y$- and $z$-directions~\footnote{The recursive equations 
for the expansion coefficients are derived from Eq.~\ref{eq:hermite-recurrence}}. 

These recursions for the expansion coefficients, along with the simple integral properties of the Hermite Gaussians, 
form the backbone of the McD scheme for integral evaluation over Gaussian basis functions.

\subsection{Hartree-Fock Equations}

Having established the Cartesian and Hermite Gaussian basis sets, we now turn to 
the matrix formulation of the HF  equations. These equations provide the foundation 
for the SCF procedure implemented in the \textsc{FSIM} library.
For pedagogical clarity, the HF equations are introduced earlier in this 
section to emphasize their direct connection to the molecular integrals developed in 
the following sections.

The objective of the HF method is to determine a set of
orthonormal molecular orbitals $\{\psi_i(\mathbf{r})\}$ that minimize the
expectation value of the electronic Hamiltonian, subject to the
constraint that the total wave function is a single Slater determinant.
Within the linear combination of atomic orbitals (LCAO) framework [see Eq.~\ref{eq:lcao}], 
each molecular orbital is expanded in the chosen Gaussian basis.
Substituting this expansion into the HF equations~\cite{SzaboOstlund1989} leads 
to a matrix eigenvalue problem known as the \emph{Roothaan equations}~\cite{Roothaan1951}:
\begin{equation}
\mathbf{F}\mathbf{C} = \mathbf{S}\mathbf{C}\boldsymbol{\varepsilon},
\label{eq:roothaan}
\end{equation}
where
 $\mathbf{F}$ is the Fock matrix,
 $\mathbf{S}$ is the overlap matrix between basis functions,
 $\mathbf{C}$ contains the molecular orbital coefficients, and
 $\boldsymbol{\varepsilon}$ is a diagonal matrix of orbital energies $\epsilon_i$.

From this point onward, we restrict ourselves to the restricted Hartree–Fock (RHF) formalism, 
in which each spatial orbital is doubly occupied by two electrons of opposite spin.
Under this framework, the electronic energy and corresponding Fock operator can be 
expressed in terms of molecular integrals.
In particular, each element of the Fock matrix is given by
\begin{equation}
F_{\mu\nu} = T_{\mu\nu} + V_{\mu\nu} + G_{\mu\nu},
\label{eq:fock-basic}
\end{equation}
which can also be expressed in the more compact form
\begin{equation}
F_{\mu\nu} =
H_{\mu\nu}^{\text{core}}
+ \sum_{\lambda\sigma} D_{\lambda\sigma}
\Big[
  (\mu\nu|\lambda\sigma)
 - \tfrac{1}{2}(\mu\lambda|\nu\sigma)
\Big],
\label{eq:fock}
\end{equation}
where $H_{\mu\nu}^{\text{core}} = T_{\mu\nu} + V_{\mu\nu}$ is the one-electron
(core) Hamiltonian matrix, and $D_{\lambda\sigma}$ is the density matrix:
\begin{equation}
D_{\lambda\sigma} = 2 \sum_{i}^{\text{occ}} C_{\lambda i} C_{\sigma i}.
\label{eq:density}
\end{equation}

The one-electron integrals appearing in Eq.~\ref{eq:fock-basic}
are defined as
\begin{align}
T_{\mu\nu} &=
\int d\mathbf{r}_1\;
\phi_{\mu}^{*}(\mathbf{r}_1)
\left[-\tfrac{1}{2}\nabla_{\mathbf{r}_1}^{2}\right]
\phi_{\nu}(\mathbf{r}_1),
\label{eq:kinetic}\\[4pt]
V_{\mu\nu} &=
\int d\mathbf{r}_1\;
\phi_{\mu}^{*}(\mathbf{r}_1)
\left[
 - \sum_A \frac{Z_A}{|\mathbf{r}_1 - \mathbf{R}_A|}
\right]
\phi_{\nu}(\mathbf{r}_1),
\label{eq:nuclear}
\end{align}
where $T_{\mu\nu}$ represents the kinetic-energy integral
and $V_{\mu\nu}$ the nuclear–attraction integral
between basis functions $\phi_{\mu}$ and $\phi_{\nu}$.

The two-electron contribution to the Fock operator,
$G_{\mu\nu}$, is expressed in terms of the
\emph{four-center electron–repulsion integrals} (ERIs):
\begin{equation}
\begin{aligned}
G_{\mu\nu}
&= \sum_{\lambda\sigma} D_{\lambda\sigma}
\Big[
  (\mu\nu|\lambda\sigma)
 - \tfrac{1}{2}(\mu\lambda|\nu\sigma)
\Big] \\[4pt]
&= \sum_{\lambda\sigma} D_{\lambda\sigma}
\Bigg[
  \iint d\mathbf{r}_1\, d\mathbf{r}_2\;
  \phi_{\mu}^{*}(\mathbf{r}_1)\phi_{\nu}(\mathbf{r}_1)
  \frac{1}{|\mathbf{r}_1 - \mathbf{r}_2|}
  \phi_{\lambda}^{*}(\mathbf{r}_2)\phi_{\sigma}(\mathbf{r}_2)
  \\[4pt]
&\hspace{2.8cm}
 - \tfrac{1}{2}
  \iint d\mathbf{r}_1\, d\mathbf{r}_2\;
  \phi_{\mu}^{*}(\mathbf{r}_1)\phi_{\lambda}(\mathbf{r}_1)
  \frac{1}{|\mathbf{r}_1 - \mathbf{r}_2|}
  \phi_{\nu}^{*}(\mathbf{r}_2)\phi_{\sigma}(\mathbf{r}_2)
\Bigg].
\end{aligned}
\label{eq:eri}
\end{equation}

The electron–repulsion integrals,
\[
(\mu\nu|\lambda\sigma)
= \iint d\mathbf{r}_1\, d\mathbf{r}_2\;
  \phi_{\mu}(\mathbf{r}_1)\phi_{\nu}(\mathbf{r}_1)
  \frac{1}{|\mathbf{r}_1 - \mathbf{r}_2|}
  \phi_{\lambda}(\mathbf{r}_2)\phi_{\sigma}(\mathbf{r}_2),
\]
are six-dimensional integrals over pairs of Gaussian basis functions and
constitute the dominant computational cost in HF calculations.
For a basis of size $N$, the formal scaling of ERI evaluation is
$\mathcal{O}(N^4)$, making their efficient computation and storage one of
the central challenges in quantum-chemical algorithms.

\subsection{Molecular Integrals}

In the \textsc{FSIM} implementation, particular attention is given to the computation 
and reuse of the molecular integrals introduced above.
These integrals constitute the core numerical workload in the self-consistent solution 
of the Roothaan equations [see Eq.~\ref{eq:roothaan}].
Their formulation and implementation in \textsc{FSIM} follow the McD 
integral scheme, which provides a systematic and recursive framework for evaluating Gaussian integrals. 
In the following subsections, we outline the mathematical formulation of each class of molecular integrals.

\subsubsection{Overlap Integrals}
\label{sec:overlap_integrals}

The overlap integrals constitute the simplest class of one-electron
integrals and appear explicitly in the HF equations
[see Eq.~\ref{eq:roothaan}]. For two primitive Cartesian Gaussians
centered on nuclei $\mathbf{A}$ and $\mathbf{B}$, the overlap integral is defined as
\begin{equation}
S_{ab} = \langle G_a | G_b \rangle
= \int G_a(\mathbf{r})\, G_b(\mathbf{r})\, d\mathbf{r},
\end{equation}
where
\begin{align}
G_a(\mathbf{r}) &=
G_{i_x i_y i_z}(\mathbf{r}; \alpha, \mathbf{A})
= G_{i_x}(\alpha,x_A)\,G_{i_y}(\alpha,y_A)\,G_{i_z}(\alpha,z_A),\\[4pt]
G_b(\mathbf{r}) &=
G_{j_x j_y j_z}(\mathbf{r}; \beta, \mathbf{B})
= G_{j_x}(\beta,x_B)\,G_{j_y}(\beta,y_B)\,G_{j_z}(\beta,z_B).
\end{align}

Because Cartesian Gaussians factorize along $x$, $y$, and $z$,
the three-dimensional overlap integral separates into a product of
one-dimensional overlaps:
\begin{equation}
S_{ab} = S_{i_x j_x}\, S_{i_y j_y}\, S_{i_z j_z},
\label{eq:s-factorization}
\end{equation}
where, for example the first term takes the form,
\begin{equation}
S_{i_x j_x} = \int G_{i_x}(\alpha,x_A)\, G_{j_x}(\beta,x_B)\, dx.
\end{equation}

To evaluate $S_{i_x j_x}$, we apply the Gaussian product theorem
and express the product of two one-dimensional Gaussians
as a single Gaussian centered at the product center $P_x$:
\begin{equation}
G_{i_x}(\alpha,x_A)\, G_{j_x}(\beta,x_B)
= \Omega_{i_x j_x}(x)
= \sum_{t=0}^{i_x + j_x} E_{t}^{i_x j_x}\, \Lambda_t(x_P),
\label{eq:omega-expansion}
\end{equation}
where $\Lambda_t(x_P)$ are the one-dimensional Hermite Gaussians and the coefficients
$E_{t}^{i_x j_x}$ are obtained recursively from
Eqs.~\ref{eq:recurrence-A}–\ref{eq:recurrence-B}.

Integrating Eq.~\ref{eq:omega-expansion} over all space yields
\begin{equation}
S_{i_x j_x} =
\int \Omega_{i_x j_x}(x)\, dx
= \sum_{t=0}^{i_x + j_x} E_{t}^{i_x j_x}
\int \Lambda_t(x_P)\, dx.
\end{equation}
Only the Hermite $s$-function ($t=0$) survives integration,
leading to
\begin{equation}
S_{i_x j_x} = E_{0}^{i_x j_x}\, \sqrt{\frac{\pi}{p}},
\qquad
p = \alpha + \beta.
\label{eq:sx-final}
\end{equation}

Combining the results for all Cartesian directions
and collecting the one-dimensional contributions,
the total three-dimensional overlap integral becomes
\begin{equation}
S_{ab}
= E_{0}^{i_x j_x}\, E_{0}^{i_y j_y}\, E_{0}^{i_z j_z}
\left( \frac{\pi}{p} \right)^{3/2}.
\label{eq:s-overall}
\end{equation}

This remarkably compact expression demonstrates how the Hermite
expansion transforms the original six-dimensional overlap integral
into a simple product of one-dimensional quantities.
All complexity is shifted to the recursive computation of the
expansion coefficients $E_{t}^{i_x j_x}$, $E_{t}^{i_y j_y}$,
and $E_{t}^{i_z j_z}$, which can be generated efficiently
using recurrence relations.

In practical implementations, the overlap matrix elements $S_{\mu\nu}$
for CGTOs are obtained by performing the linear
contractions over primitives, as described in the previous sections.
These integrals form a key component of the Roothaan--Hall equations
and enter directly in the orthogonalization of the molecular orbital
basis and in the construction of the Fock matrix.

\subsubsection{Kinetic–Energy Integrals}

The kinetic–energy matrix elements, which enter the one-electron part
of the Fock operator [see Eq.~\ref{eq:fock-basic}], can also be
expressed in terms of overlap integrals over Cartesian Gaussians.
For two primitive Cartesian Gaussians centered at $\mathbf{A}$ and
$\mathbf{B}$ with exponents $\alpha$ and $\beta$, respectively, the
kinetic–energy integral is defined as
\begin{equation}
\mathcal{T}_{ab}
= -\frac{1}{2}\,
\big\langle G_a \,\big|\,
\nabla^2
\,\big|\, G_b \big\rangle
= -\frac{1}{2}\,
\big\langle G_a \,\big|\,
\partial_x^2 + \partial_y^2 + \partial_z^2
\,\big|\, G_b \big\rangle .
\label{eq:Tab-def}
\end{equation}

Each primitive Gaussian is separable along the Cartesian directions,
\[
G_a(\mathbf{r}) =
G_{i_x}(\alpha, x_A)\, G_{i_y}(\alpha, y_A)\, G_{i_z}(\alpha, z_A),
\qquad
G_b(\mathbf{r}) =
G_{j_x}(\beta, x_B)\, G_{j_y}(\beta, y_B)\, G_{j_z}(\beta, z_B),
\]
allowing the kinetic–energy integral to factorize as
\begin{equation}
\mathcal{T}_{ab}
= \mathcal{T}_{i_x j_x}\, S_{i_y j_y}\, S_{i_z j_z}
+ S_{i_x j_x}\, \mathcal{T}_{i_y j_y}\, S_{i_z j_z}
+ S_{i_x j_x}\, S_{i_y j_y}\, \mathcal{T}_{i_z j_z},
\label{eq:Tab-factor}
\end{equation}
where $S_{i_x j_x}$ and $T_{i_x j_x}$ denote, respectively,
the one-dimensional overlap and kinetic–energy matrix elements:
\[
S_{i_x j_x} = \langle G_{i_x}(\alpha, x_A) \mid G_{j_x}(\beta, x_B) \rangle,
\qquad
\mathcal{T}_{i_x j_x} =
-\tfrac{1}{2}\,
\langle G_{i_x}(\alpha, x_A)
\mid \partial_x^2
\mid G_{j_x}(\beta, x_B) \rangle.
\]

To derive the kinetic integral, we start from the second derivative of a
Cartesian Gaussian with respect to its center coordinate. For the $x$
component:
\begin{align}
\frac{d}{dx}\,G_{j_x}(\beta,x_B)
&= -2\beta\, G_{j_x+1}(\beta,x_B)
   + j_x\, G_{j_x-1}(\beta,x_B), \\[4pt]
\frac{d^2}{dx^2}\,G_{j_x}(\beta,x_B)
&= 4\beta^2\, G_{j_x+2}(\beta,x_B)
 - 2\beta(2j_x+1)\, G_{j_x}(\beta,x_B)
 + j_x(j_x-1)\, G_{j_x-2}(\beta,x_B).
\label{eq:second-derivative}
\end{align}

Substituting Eq.~\ref{eq:second-derivative} into the definition of
$\mathcal{T}_{i_x j_x}$ and recognizing the resulting overlap integrals yields
\begin{equation}
\mathcal{T}_{i_x j_x}
= -2\beta^2\, S_{i_x, j_x+2}
 + \beta(2j_x+1)\, S_{i_x j_x}
 - \tfrac{1}{2}\, j_x(j_x-1)\, S_{i_x, j_x-2}.
\label{eq:Tij-from-S}
\end{equation}

Using the Hermite expansion of the overlaps introduced in
Section~\ref{sec:overlap_integrals}, the integrals in Eq.~\ref{eq:Tij-from-S} can be expressed
in terms of the Hermite expansion coefficients $E_{t}^{i_x j_x}$.
Since only the Hermite $s$-function ($t=0$) contributes upon integration,
we obtain
\begin{equation}
\mathcal{T}_{i_x j_x}
=\Big[
 -2\beta^2 E_{i_x, j_x+2}^{0}
 + \beta(2j_x+1) E_{i_x j_x}^{0}
 - \tfrac{1}{2} j_x(j_x-1) E_{i_x, j_x-2}^{0}
\Big]
\sqrt{\frac{\pi}{p}},
\qquad p = \alpha + \beta.
\label{eq:Tij-hermite}
\end{equation}

Combining the three Cartesian contributions
and collecting the one-dimensional results,
the complete three-dimensional kinetic–energy integral can be written as
\begin{equation}
\begin{aligned}
\mathcal{T}_{ab} &=
\Big[
 -2\beta^2 E_{i_x, j_x+2}^{0}
 + \beta(2j_x+1)E_{i_x j_x}^{0}
 - \tfrac{1}{2}j_x(j_x-1)E_{i_x, j_x-2}^{0}
\Big]
E_{i_y j_y}^{0}\, E_{i_z j_z}^{0}
\left( \frac{\pi}{p} \right)^{3/2} \\[6pt]
&\quad +
\Big[
 -2\beta^2 E_{i_y, j_y+2}^{0}
 + \beta(2j_y+1)E_{i_y j_y}^{0}
 - \tfrac{1}{2}j_y(j_y-1)E_{i_y, j_y-2}^{0}
\Big]
E_{i_x j_x}^{0}\, E_{i_z j_z}^{0}
\left( \frac{\pi}{p} \right)^{3/2} \\[6pt]
&\quad +
\Big[
 -2\beta^2 E_{i_z, j_z+2}^{0}
 + \beta(2j_z+1)E_{i_z j_z}^{0}
 - \tfrac{1}{2}j_z(j_z-1)E_{i_z, j_z-2}^{0}
\Big]
E_{i_x j_x}^{0}\, E_{i_y j_y}^{0}
\left( \frac{\pi}{p} \right)^{3/2}.
\end{aligned}
\label{eq:Tab-final}
\end{equation}

In the case CGTOs, which serve as the basis 
functions in the Roothaan formulation, the corresponding matrix 
elements are computed as linear combinations of the primitive contributions weighted 
by the contraction coefficients:
\begin{equation}
T_{\mu\nu}
= \sum_{k}^{K_\mu}
  \sum_{l}^{K_\nu}
  d_{k}^{\mu} d_{l}^{\nu}\,
  \mathcal{T}_{kl},
\end{equation}
where $K_\mu$ and $K_\nu$ denote the \emph{degrees of contraction}, that is, the numbers of primitive 
functions in the CGTOs $\phi_\mu$ and $\phi_\nu$, respectively.

These kinetic–energy integrals, together with the 
nuclear–attraction integrals, form the core one-electron part of the
HF Hamiltonian. Their analytical tractability and recursive structure make them
particularly suitable for efficient implementation within the
\textsc{FSIM} integral engine.

\subsubsection{Two–Center One–Electron Integrals}

The second class of one–electron integrals contributing to
$H^{\text{core}}$ are the \emph{electron–nucleus attraction integrals},
which describe the Coulomb interaction between an electron and the
nuclear charge distribution.
For two primitive Cartesian Gaussians centered at $\mathbf{A}$ and
$\mathbf{B}$, and a nucleus located at $\mathbf{C}$, these integrals are defined as
\begin{equation}
\mathcal{V}_{ab}
= \left\langle G_a \,\middle|\, r_{C}^{-1} \,\middle|\, G_b \right\rangle
= \int \frac{G_a(\mathbf{r})\, G_b(\mathbf{r})}{|\mathbf{r}-\mathbf{C}|}\,
d\mathbf{r}.
\label{eq:Vab-def}
\end{equation}

Using the Gaussian product theorem
and the Hermite expansion introduced previously
[see Eq.~\ref{eq:omega-expansion}],
the product of two Cartesian Gaussians is expressed as
\begin{equation}
G_a(\mathbf{r})\, G_b(\mathbf{r})
= \sum_{tuv} E_{t}^{i_x j_x}\, E_{u}^{i_y j_y}\, E_{v}^{i_z j_z}\,
\Lambda_{tuv}(\mathbf{r}_P; p, \mathbf{P}),
\label{eq:Vab-hermite-expansion}
\end{equation}
where $p = \alpha + \beta$ is the combined exponent and
$\mathbf{P}$ is the product center
$\mathbf{P} = (\alpha\mathbf{A} + \beta\mathbf{B})/p$.
Substituting Eq.~\ref{eq:Vab-hermite-expansion}
into Eq.~\ref{eq:Vab-def} yields
\begin{equation}
\mathcal{V}_{ab}
= \sum_{tuv}
E_{t}^{i_x j_x}\, E_{u}^{i_y j_y}\, E_{v}^{i_z j_z}
\int \frac{\Lambda_{tuv}(\mathbf{r}_{P})}{|\mathbf{r}-\mathbf{C}|}\,
d\mathbf{r}.
\label{eq:Vab-hermite-int}
\end{equation}

The integral in Eq.~\ref{eq:Vab-hermite-int} defines \emph{the Hermite Coulomb integral} or
auxiliary Hermite integral of order zero,
\begin{equation}
\int \frac{\Lambda_{tuv}(\mathbf{r}_{P})}{|\mathbf{r}-\mathbf{C}|}\, d\mathbf{r}
= \frac{2\pi}{p}\,
R_{tuv}^{0}(p,\mathbf{R}_{PC}),
\label{eq:R0-definition}
\end{equation}
where $\mathbf{R}_{PC} = \mathbf{P}-\mathbf{C}$ is the distance vector
from the product center to the nucleus~$C$.
Thus, the electron–nucleus integral becomes
\begin{equation}
\mathcal{V}_{ab}
= \frac{2\pi}{p}
\sum_{tuv}
E_{t}^{i_x j_x}\, E_{u}^{i_y j_y}\, E_{v}^{i_z j_z}\,
R_{tuv}^{0}(p,\mathbf{R}_{PC}).
\label{eq:Vab-final}
\end{equation}

The auxiliary Hermite integrals of arbitrary order, $R_{tuv}^{n}$,
are expressed in terms of the \emph{Boys function} $F_{n}(x)$ as
\begin{equation}
R_{tuv}^{n}(p,\mathbf{R}_{PC})
= (-2p)^{n}
\frac{\partial^{t+u+v}}
{\partial P_{x}^{t}\, \partial P_{y}^{u}\, \partial P_{z}^{v}}
F_{n}\!\left(p\, R_{PC}^{2}\right),
\label{eq:Rn-definition}
\end{equation}
where
\[
F_{n}(x)
= \int_{0}^{1} e^{-x t^{2}}\, t^{2n}\, dt
\]
is the standard Boys function.

In practice, the auxiliary Hermite integrals
$R_{tuv}^{n}$ are generated recursively from the source term
\[
R_{000}^{n} = (-2p)^{n} F_{n}\!\left(p R_{PC}^{2}\right),
\]
using the McD recurrence relations \cite{Helgaker2000, MCMURCHIE1978218}:
\begin{equation}
\begin{aligned}
R_{t+1,\,u,\,v}^{n} &= t\, R_{t-1,\,u,\,v}^{\,n+1}
                     + X_{PC}\, R_{tuv}^{\,n+1}, \\[4pt]
R_{t,\,u+1,\,v}^{n} &= u\, R_{t,\,u-1,\,v}^{\,n+1}
                     + Y_{PC}\, R_{tuv}^{\,n+1}, \\[4pt]
R_{t,\,u,\,v+1}^{n} &= v\, R_{t,\,u,\,v-1}^{\,n+1}
                     + Z_{PC}\, R_{tuv}^{\,n+1}.
\end{aligned}
\label{eq:R-recursion}
\end{equation}

These relations are applied recursively until all desired
$R_{tuv}^{0}$ terms have been obtained.
In combination with Eq.~\ref{eq:Vab-final},
they provide a complete and efficient analytical scheme for
evaluating the electron–nucleus attraction integrals.


In the case of CGTOs, the corresponding matrix elements of the 
electron–nucleus potential are obtained as weighted sums of the primitive contributions, 
using the contraction coefficients as weights:
\begin{equation}
V_{\mu\nu}
= \sum_{k}^{K_\mu}
  \sum_{l}^{K_\nu}
  d_{k}^{\mu} d_{l}^{\nu}\,
  \mathcal{V}_{kl},
\end{equation}
where $K_\mu$ and $K_\nu$ denote the number of primitives in
the CGTOs $\phi_\mu$ and $\phi_\nu$.

The analytical form of Eqs.~(\ref{eq:R0-definition})–(\ref{eq:R-recursion})
allows these integrals to be computed with high numerical stability and
efficiency.

\subsubsection{Two–Electron Repulsion Integrals}

The two–electron repulsion integrals (ERIs), which describe the Coulomb
interaction between pairs of electrons, constitute the most computationally
demanding part of the HF method. Within the McD
scheme, their evaluation follows the same general procedure as for the
one–electron integrals, employing Hermite expansions and auxiliary Hermite
integrals.

For four primitive Cartesian Gaussians centered at
$\mathbf{A}$, $\mathbf{B}$, $\mathbf{C}$, and $\mathbf{D}$ with exponents
$\alpha$, $\beta$, $\gamma$, and $\delta$, respectively, the basic integral is
defined as \cite{SzaboOstlund1989, Helgaker2000}
\begin{equation}
[ab|cd]
= \iint
  \frac{G_a(\mathbf{r}_1)\, G_b(\mathbf{r}_1)\,
        G_c(\mathbf{r}_2)\, G_d(\mathbf{r}_2)}
       {|\mathbf{r}_1 - \mathbf{r}_2|}
  \, d\mathbf{r}_1\, d\mathbf{r}_2,
\label{eq:twoe-basic}
\end{equation}
where each $G$ denotes a primitive Cartesian Gaussian–type orbital (CGTO).


The product of two Gaussians on the same electron can be expanded as an
\emph{overlap distribution} centered at the corresponding product center.
Using the notation introduced previously,
\begin{align}
\Omega_{ab}(\mathbf{r}_1)
&= G_a(\mathbf{r}_1)\, G_b(\mathbf{r}_1)
 = \sum_{tuv} E_{t}^{i_x j_x}\, E_{u}^{i_y j_y}\, E_{v}^{i_z j_z}\,
   \Lambda_{tuv}(\mathbf{r}_1 - \mathbf{P}; p, \mathbf{P}), \\[4pt]
\Omega_{cd}(\mathbf{r}_2)
&= G_c(\mathbf{r}_2)\, G_d(\mathbf{r}_2)
 = \sum_{\tau\nu\phi}
   E_{\tau}^{k_x l_x}\, E_{\nu}^{k_y l_y}\, E_{\phi}^{k_z l_z}\,
   \Lambda_{\tau\nu\phi}(\mathbf{r}_2 - \mathbf{Q}; q, \mathbf{Q}),
\end{align}
where $p = \alpha + \beta$, $q = \gamma + \delta$, and
$\mathbf{P}$ and $\mathbf{Q}$ are the respective product centers:
\[
\mathbf{P} = \frac{\alpha \mathbf{A} + \beta \mathbf{B}}{p},
\qquad
\mathbf{Q} = \frac{\gamma \mathbf{C} + \delta \mathbf{D}}{q}.
\]

Substituting these expansions into Eq.~\ref{eq:twoe-basic} gives
\begin{equation}
[ab|cd]
= \sum_{tuv} E_{t}^{i_x j_x}\, E_{u}^{i_y j_y}\, E_{v}^{i_z j_z}
  \sum_{\tau\nu\phi}
  E_{\tau}^{k_x l_x}\, E_{\nu}^{k_y l_y}\, E_{\phi}^{k_z l_z}
  \iint
  \frac{\Lambda_{tuv}(\mathbf{r}_1; p, \mathbf{P})\,
        \Lambda_{\tau\nu\phi}(\mathbf{r}_2; q, \mathbf{Q})}
       {|\mathbf{r}_1 - \mathbf{r}_2|}
  \, d\mathbf{r}_1\, d\mathbf{r}_2.
\label{eq:twoe-expanded}
\end{equation}


The remaining two–electron integral in Eq.~\ref{eq:twoe-expanded}
involves the Coulomb interaction between two Hermite Gaussians,
one centered at $\mathbf{P}$ and the other at $\mathbf{Q}$.
Following the derivation in Ref.~\cite{Helgaker2000}, this integral can be evaluated analytically as
\begin{align}
\iint
\frac{\Lambda_{tuv}(\mathbf{r}_1; p, \mathbf{P})\,
      \Lambda_{\tau\nu\phi}(\mathbf{r}_2; q, \mathbf{Q})}
     {|\mathbf{r}_1 - \mathbf{r}_2|}
\, d\mathbf{r}_1\, d\mathbf{r}_2
&= (-1)^{\tau+\nu+\phi}
   \frac{2 \pi^{5/2}}{p q \sqrt{p+q}} \notag \\[6pt]
&\quad \times
\left( \frac{\partial}{\partial P_x} \right)^{t+\tau}
\left( \frac{\partial}{\partial P_y} \right)^{u+\nu}
\left( \frac{\partial}{\partial P_z} \right)^{v+\phi}
F_{0}\!\left(\alpha R_{PQ}^{2}\right),
\label{eq:twoe-derivative}
\end{align}
where $\alpha = \tfrac{p q}{p + q}$,
$\mathbf{R}_{PQ} = \mathbf{P} - \mathbf{Q}$,
and $F_0$ is the zeroth–order Boys function.


The expression above can be written compactly in terms of the Hermite Coulomb integrals
$R^{0}_{tuv}$ introduced earlier:
\begin{equation}
\iint
\frac{\Lambda_{tuv}(\mathbf{r}_1)\,
      \Lambda_{\tau\nu\phi}(\mathbf{r}_2)}
     {|\mathbf{r}_1 - \mathbf{r}_2|}
\, d\mathbf{r}_1\, d\mathbf{r}_2
= (-1)^{\tau+\nu+\phi}
  \frac{2 \pi^{5/2}}{p q \sqrt{p+q}}\,
  R^{0}_{t+\tau,\,u+\nu,\,v+\phi}
  \!\left(\alpha, \mathbf{R}_{PQ}\right).
\label{eq:twoe-R}
\end{equation}

Substituting Eq.~\ref{eq:twoe-R} into Eq.~\ref{eq:twoe-expanded}
yields the final form of the two–electron repulsion integral:
\begin{equation}
[ab|cd]
= \frac{2 \pi^{5/2}}{p q \sqrt{p+q}}
  \sum_{tuv} \sum_{\tau\nu\phi}
  (-1)^{\tau+\nu+\phi}
  E_{t}^{i_x j_x} E_{u}^{i_y j_y} E_{v}^{i_z j_z}
  E_{\tau}^{k_x l_x} E_{\nu}^{k_y l_y} E_{\phi}^{k_z l_z}
  R^{0}_{t+\tau,\,u+\nu,\,v+\phi}
  \!\left(\alpha, \mathbf{R}_{PQ}\right).
\label{eq:twoe-final}
\end{equation}


In the Roothaan formulation, the CGTOs
used to build the Fock matrix are linear combinations of
primitives. Consequently, the two–electron matrix elements are obtained
as weighted sums over the primitive ERIs:
\begin{equation}
(\phi_\mu \phi_\nu| \phi_\lambda \phi_\sigma)
= \sum_{k}^{K_\mu}
  \sum_{l}^{K_\nu}
  \sum_{m}^{K_\lambda}
  \sum_{n}^{K_\sigma}
  d_{k}^{\mu} d_{l}^{\nu} d_{m}^{\lambda} d_{n}^{\sigma}\,
  [a_k b_l|c_m d_n].
\end{equation}

These four–center integrals constitute the electron–electron term in the
Fock matrix [see Eq.~\ref{eq:fock}], and dominate the computational cost
of HF and post–Hartree–Fock methods. Within the
\textsc{FSIM} framework, the evaluation of these integrals leverages the
recursive structure of the auxiliary Hermite integral relations and optimized
schemes to balance accuracy and efficiency.

\subsection{Contracted Shells, Shell Pairs, and Shell Quartets}

In order to understand the terminology used in algorithms for the computation 
of molecular integrals, it is useful to introduce the concepts of contracted shells, shell pairs, and shell quartets.
The introduction of these constructs exploits the fact that basis functions sharing a common center 
and exponent set—originally designed to resemble atomic orbitals—allow extensive reuse of intermediate 
quantities, thereby greatly simplifying and accelerating the evaluation of Gaussian integrals.

The set of all CGTFs sharing the same center and the 
same set of primitive exponents constitutes a \emph{contracted shell}~\cite{Peter1997}.  
For example, the three functions $\{p_x, p_y, p_z\}$ on a given center form one contracted $p$-shell.

In the McMurchie--Davidson scheme, integrals are expressed in terms of products of basis functions.  
The product of two CGTFs on centers $\mathbf{A}$ and $\mathbf{B}$,
\[
(ab| = | \phi_a(\mathbf{r}_1)\phi_b(\mathbf{r}_1) ),
\]
is associated with a pair of shells on centers $A$ and $B$ and is called a 
\emph{contracted shell pair}.  
Analogously, the product
\[
|cd) = | \phi_c(\mathbf{r}_2)\phi_d(\mathbf{r}_2) )
\]
defines another contracted shell pair on centers $C$ and $D$.

The complete four-center product appearing in the two-electron integral $(ab|cd)$
is termed a \emph{contracted shell quartet}.  
The bra $(ab|$ and the ket $|cd)$ correspond to the two contracted shell pairs that define the integral.  
The total degree of contraction for the integral is 
\[
K_{\mathrm{tot}} = (K_a K_b)(K_c K_d),
\]
and its total angular momentum is 
\[
L_{\mathrm{tot}} = l_a + l_b + l_c + l_d.
\]

These parameters are easily established for a given contracted shell quartet and 
provide valuable insight into the computational complexity associated with evaluating 
the corresponding two-electron integrals.

\subsection{Contraction of Two–Electron Integrals}

The four–index electron–repulsion integrals (ERIs), which describe the
Coulomb interaction between pairs of electrons, form the most
computationally demanding tensor in the HF method. For a system
with $N$ basis functions, the number of unique integrals scales formally
as $\mathcal{O}(N^{4})$, and their efficient evaluation and contraction
are critical in all electronic–structure implementations.

Following the McD scheme, contracted ERIs are expressed as linear combinations of primitive Cartesian integrals, whose overlap distributions are expanded in terms of Hermite Gaussians, yielding:
\begin{equation}
\begin{aligned}
(\phi_a \phi_b \mid \phi_c \phi_d)
&= \sum_{\substack{
     k_a,~k_b,\\
     k_c,~k_d
   }}
   d_{k_a}\, d_{k_b}\, d_{k_c}\, d_{k_d} \\[6pt]
&\quad \times
   \sum_{\substack{
     i_x i_y i_z,\\
     j_x j_y j_z,\\
     k_x k_y k_z,\\
     l_x l_y l_z
   }}
   S_{i_x i_y i_z}^{l_a m_a}\,
   S_{j_x j_y j_z}^{l_b m_b}\,
   S_{k_x k_y k_z}^{l_c m_c}\,
   S_{l_x l_y l_z}^{l_d m_d} \\[6pt]
&\quad \times
   \sum_{\substack{
     t u v,\\
     \tau \lambda \kappa
   }}
   E_{t}^{i_x j_x}\, E_{u}^{i_y j_y}\, E_{v}^{i_z j_z}\,
   E_{\tau}^{k_x l_x}\, E_{\lambda}^{k_y l_y}\, E_{\kappa}^{k_z l_z}\,
   R^{0}_{t+\tau,\,u+\lambda,\,v+\kappa}
   (\alpha_{\boldsymbol{k}}, \mathbf{R}_{PQ})
   \Biggr\}.
\end{aligned}
\label{eq:eri-tensor-hermite}
\end{equation}

Here, $\{t, u, v, \tau, \lambda, \kappa\}$ denote Hermite indices
arising from the expansion of Gaussian products, and
$R^{0}_{t+\tau,\,u+\lambda,\,v+\kappa}$ are the Hermite Coulomb integrals
defined in Eq.~\ref{eq:R0-definition}. We have introduced the index $\boldsymbol{k} = \{k_a, k_b, k_c, k_d \}$ to 
make explicit the  dependence of the Hermite integrals on the configuration of contraction parameters. 

At the level of primitive integrals, several Cartesian index combinations
map onto the same Hermite triplet $(t,u,v)$ for a given shell pair.
It is therefore advantageous to contract these redundant terms prior to
the evaluation of the auxiliary Hermite integrals. The contraction gathers
all Cartesian components corresponding to identical Hermite orders into a
single intermediate tensor:
\begin{equation}
E_{tuv}^{k_a k_b}
= \sum_{i_x i_y i_z}\sum_{j_x j_y j_z}
  S_{i_x i_y i_z}^{l_a m_a}\,
  S_{j_x j_y j_z}^{l_b m_b}\,
  E_{t}^{i_x j_x}\, E_{u}^{i_y j_y}\, E_{v}^{i_z j_z}.
\label{eq:E-tuv}
\end{equation}
An analogous contraction is applied to the second electron pair,
yielding $E_{\tau\lambda\kappa}^{k_c k_d}$. The ERI tensor then takes the
compact contracted form
\begin{equation}
(\phi_a \phi_b \mid \phi_c \phi_d)
=  \sum_{\substack{
     k_a,~k_b,\\
     k_c,~k_d
  }}
  d_{k_a} d_{k_b} d_{k_c} d_{k_d}
  \sum_{\substack{
     t u v,\\
     \tau \lambda \kappa
  }}
  E_{tuv}^{k_a k_b}\,
  E_{\tau\lambda\kappa}^{k_c k_d}\,
  R^{0}_{t+\tau,\,u+\lambda,\,v+\kappa}
  (\alpha_{\boldsymbol{k}}, \mathbf{R}_{PQ}).
\label{eq:eri-tensor-contracted}
\end{equation}

This contraction over Cartesian indices reduces the number of intermediate
quantities and improves data locality and numerical efficiency. The resulting
expressions form the computational kernel of the McD and
related integral–reduction schemes used in modern quantum–chemistry software.

Another possible intermediate contraction produces the following tensor:
\begin{equation}
J_{t,u,v, \boldsymbol{k}}
=  \sum_{\tau \lambda \kappa}
  E_{\tau\lambda\kappa}^{k_c k_d}\,
  R^{0}_{t+\tau,\,u+\lambda,\,v+\kappa}
  (\alpha_{\boldsymbol{k}}, \mathbf{R}_{PQ}).
\end{equation}
which can reduce the computational complexity of Eq.~\ref{eq:eri-tensor-contracted}. 
See Ref.~\cite{Helgaker2000} for an extended discussion of contractions and their implications for the computational complexity of ERIs.


\begin{figure}[!htbp]
  \centering
  \includegraphics[width=0.92\textwidth]{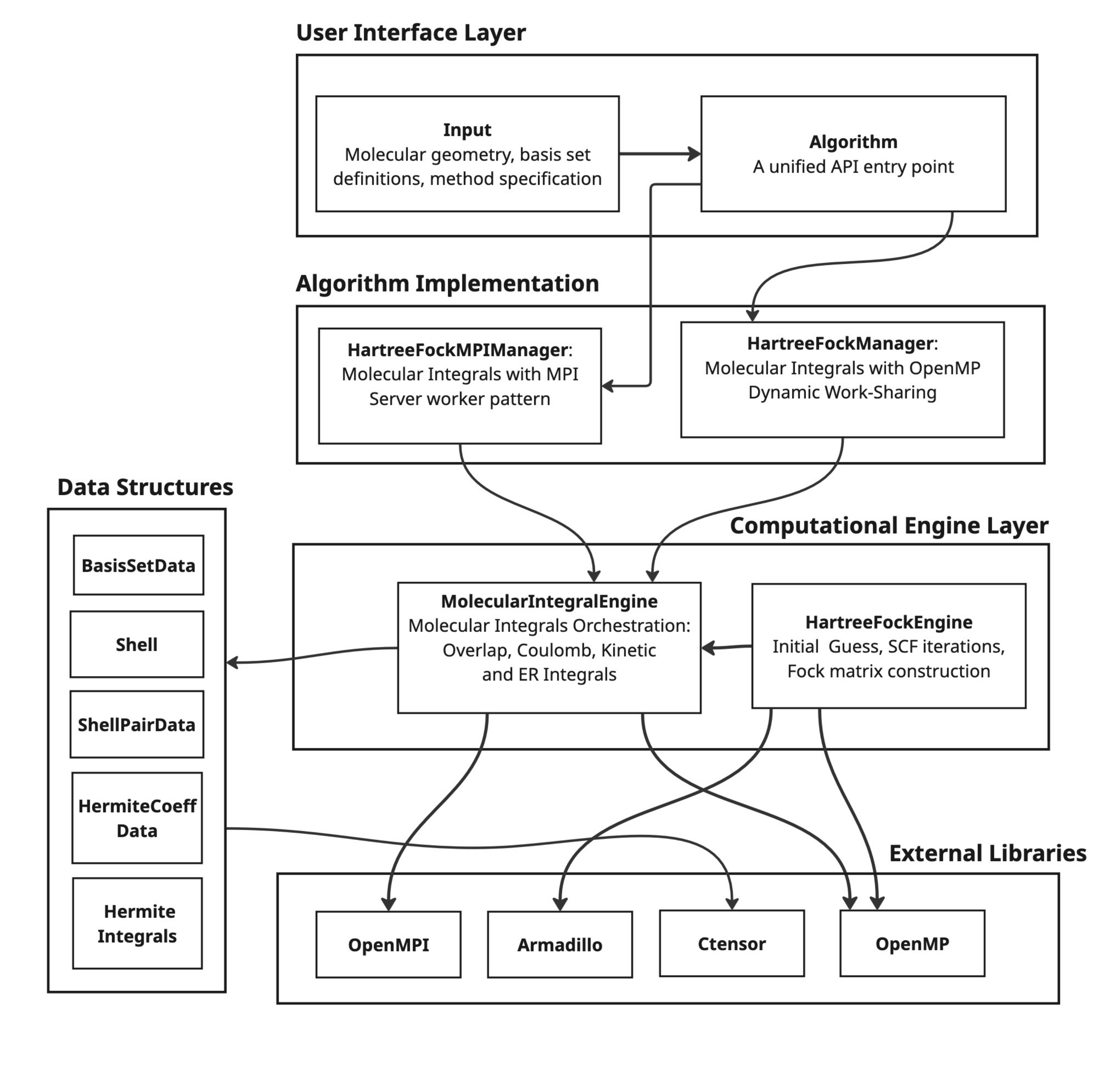}
  \caption{Layered architecture of the FSIM library. The system is organized into three 
  main layers: the \emph{User Interface Layer}, which handles user input and provides 
  a unified API through the \texttt{Algorithm} class; the \emph{Algorithm Implementation Layer}, which 
  coordinates the execution of Hartree-Fock calculations using either OpenMP or MPI parallel 
  strategies; and the \emph{Computational Engine Layer}, which performs molecular integral 
  evaluation and SCF iterations through the \texttt{MolecularIntegralEngine} and \texttt{HartreeFockEngine} 
  components. Core data structures such as \texttt{BasisSetData}, \texttt{ShellPairData}, and \texttt{HermiteCoeffData} support 
  the numerical computations, while external dependencies (\texttt{OpenMPI}, \texttt{Armadillo}, \texttt{Ctensor}, and \texttt{OpenMP}) 
  provide linear algebra, tensor–based data structures, and parallelization capabilities. To maintain simplicity, only the main classes 
  are shown, omitting lower-level implementation details. 
  The source code is hosted here: \href{https://gitlab.com/marhvera/fsim}{https://gitlab.com/marhvera/fsim}.}
  \label{fig:hf_flow}
\end{figure}

\section{FSIM: Software Design and Implementation}
\label{sec:implementation}

\subsection{Architecture Overview}
FSIM is a C++ library intended for instructional and research-training purposes, 
focused on illustrating the computation of Hartree–Fock (HF) molecular energies with 
transparent, reproducible algorithms.
It combines a modular software design with 
well-established theoretical foundations, particularly the McD method for 
evaluating molecular integrals and a SCF procedure for electronic 
structure optimization. FSIM is implemented with parallel capabilities using both OpenMP and 
MPI, allowing execution on shared-memory systems and, in principle, distributed-memory environments.

At its core, the library features a clean, layered architecture that separates the responsibilities 
of input parsing, algorithmic coordination, and numerical computation, as illustrated 
in Fig.\ref{fig:hf_flow}. The user interacts with the library primarily through 
the high-level \texttt{fsim::Algorithm} class, which serves as the main entry point 
for performing electronic structure calculations. A typical workflow involves preparing 
an input file that specifies the computational method, molecular geometry, and Gaussian 
basis set; constructing an \texttt{Algorithm} object; invoking the desired calculation 
through instances of \texttt{AbstractManager} (such as \texttt{HartreeFockManager}); and retrieving 
the results—such as the total molecular energy—from the \texttt{ManagerResult} structure. This minimal 
interface enables users to perform standard HF calculations with ease, while retaining 
full control over the computational environment and parallelization settings. An example code snippet is shown below:

\vspace{0.2 cm }

\begin{lstlisting}[caption={Example of user code for computing the HF energy with FSIM and OpenMP.}, label={lst:fsim-example}]
#include <fsim.h>
#include <string>
#include <iostream>

int main() {
  using namespace fsim;
  using namespace std;
  string file_name = "lih_aug-cc-pvdz.txt";
  Algorithm algo(file_name);
  algo.run("hartree_fock");
  cout << "hf_energy = " << algo.result.hf_mol_energy << '\n';
}
\end{lstlisting}

\begin{lstlisting}[caption={Example of user code for computing the HF energy with FSIM and MPI.}, label={lst:fsim-example}]
#include <fsim.h>
#include <string>
#include <iostream>
#include <mpi.h>

int main(int argc, char* argv[]) {
  using namespace fsim;
  using namespace std;

  MPI_Init(&argc, &argv);

  int rank;
  MPI_Comm_rank(MPI_COMM_WORLD, &rank);

  string file_name = "lih_aug-cc-pvdz.txt";
  Algorithm algo(file_name, argc, argv);
  algo.run("hartree_fock_mpi");

  if (rank == 0) {
    cout << "hf_energy = " << algo.result.hf_mol_energy << '\n';
  }

  MPI_Finalize();
}
\end{lstlisting}

Internally, FSIM is organized into three conceptual layers that reflect 
the logical structure of a quantum chemistry code. The \emph{input processing layer} handles 
all aspects of reading and interpreting the user input, transforming molecular data, basis 
set definitions, and charge and spin information into structured objects such as \texttt{Molecule}, \texttt{BasisSetData}, 
and \texttt{ShellPairData}. These data structures provide the foundation for subsequent 
computational steps. The \emph{algorithm layer} orchestrates the overall workflow, implementing 
the HF managers---both in sequential and parallel forms---through a common abstract 
interface. This design enables the same client code to execute on different hardware configurations 
or algorithmic variants without modification. The \emph{computational engine layer} carries out 
the numerically intensive operations, including integral generation and SCF iterations, through 
specialized classes such as \texttt{MolecularIntegralEngine} for molecular integrals 
and \texttt{HartreeFockEngine} for iterative diagonalization and density matrix updates.

Through its object-oriented design, FSIM maintains clear ownership and data flow between 
its layers: molecular and basis data are produced by the input system, integral tensors 
are computed by the engine, and results are aggregated at the algorithm level and returned 
to the user through a uniform interface. This separation of concerns simplifies testing, 
enables reuse of computational components, and provides a clean foundation for future extensions 
such as GPU acceleration.

Finally, FSIM provides an extensible and developer-friendly framework. Its CMake-based build 
system, containerized development environment, and clear modular boundaries make it suitable 
for experimentation and development of new features. The library thus serves as a practical and 
pedagogical platform for exploring the algorithms, new data structures, and performance strategies 
underlying modern \emph{ab initio} quantum chemistry.

\subsection{SCF Solver and DIIS Acceleration}

In the FSIM library, the HF method is implemented as a self-consistent 
field (SCF) procedure that iteratively determines the molecular electronic structure by 
solving the Roothaan equations in a Gaussian basis. The objective is to obtain a self-consistent 
density matrix $\mathbf{D}$ such that the Fock operator $\mathbf{F}[\mathbf{D}]$ generates molecular 
orbitals reproducing the same density. The implementation adheres to the canonical HF workflow but 
adopts a modular architecture that integrates seamlessly with the McMurchie–Davidson integral 
engine and FSIM’s algorithm orchestration framework.

The entire SCF process is managed by the \texttt{HartreeFockEngine} class, which 
encapsulates all key components of the iteration—construction of the Fock and density 
matrices, computation of the G-matrix, diagonalization, and convergence control. It serves 
as the top-level driver, coordinating precomputed integrals from the \texttt{MolecularIntegralEngine} 
and executing the SCF cycle through the \texttt{drive\_scf()} interface. A typical HF calculation 
in FSIM proceeds as follows:

\begin{enumerate}
  \item \textbf{Initial Guess for the Density Matrix:}  
  The function \texttt{GuessDensityMatrix()} generates an initial density matrix $\mathbf{D}^{(0)}$.  
  In FSIM, this is done by setting the Fock matrix to the core Hamiltonian $\mathbf{H}^{\text{core}}$ and diagonalizing it in the orthogonal basis, yielding initial molecular orbital coefficients and occupancies.

  \item \textbf{Compute the One-Electron Matrices:}  
  The McMurchie--Davidson module provides the one-electron integrals that form the core Hamiltonian,
  \[
  \mathbf{H}^{\text{core}} = \mathbf{T} + \mathbf{V},
  \]
  where $\mathbf{T}$ and $\mathbf{V}$ are the kinetic and nuclear-attraction matrices, respectively.  
  The overlap matrix $\mathbf{S}$ is also computed during this stage and validated for Hermiticity.

  \item \textbf{Symmetric Orthogonalization of the Basis:}  
  The basis set is transformed into an orthonormal representation using the symmetric orthogonalization matrix $\mathbf{X}$,
  \[
  \mathbf{X} = \mathbf{U}\,\mathbf{s}^{-1/2}\,\mathbf{U}^{\dagger},
  \]
  where $\mathbf{U}$ and $\mathbf{s}$ are the eigenvectors and eigenvalues of the overlap matrix $\mathbf{S}$.  
  This transformation, implemented in \texttt{ComputeXRotMatrix()}, ensures that the generalized eigenvalue problem becomes standard in the orthogonal basis.

  \item \textbf{Iterative SCF Cycle:}
  \begin{enumerate}
    \item \textbf{Build the G-Matrix:}  
    The two-electron (Coulomb and exchange) contributions are assembled using the expression
    \[
    G_{\mu\nu} = \sum_{\lambda\sigma} D_{\lambda\sigma} 
    \big[(\mu\nu|\sigma\lambda) - \tfrac{1}{2}(\mu\lambda|\sigma\nu)\big],
    \]
    implemented in \texttt{ComputeGFockMatrix()}.  
    This step is the most computationally demanding part of the HF algorithm and dominates runtime with an $O(N^4)$ scaling, where $N$ is the number of basis functions.  
    FSIM employs OpenMP parallelization with
    \texttt{\#pragma omp parallel for collapse(2) schedule(dynamic)}  
    to distribute this workload across threads efficiently.

    \item \textbf{Assemble the Fock Matrix:}  
    The one-electron and two-electron terms are combined to form the Fock operator,
    \[
    \mathbf{F} = \mathbf{H}^{\text{core}} + \mathbf{G}(\mathbf{D}),
    \]
    implemented in \texttt{ComputeFockMatrix()}.

    \item \textbf{Transform and Diagonalize the Fock Matrix:}  
    The Fock matrix is transformed to the orthogonal basis:
    \[
    \mathbf{F}' = \mathbf{X}^{\dagger} \mathbf{F} \mathbf{X},
    \]
    and diagonalized using \texttt{DiagonalizeFockMatrix()} to obtain orbital energies $\boldsymbol{\epsilon}$ and coefficients $\mathbf{C}'$.  
    The coefficients are then back-transformed to the atomic basis as $\mathbf{C} = \mathbf{X}\,\mathbf{C}'$.

    \item \textbf{Update the Density Matrix:}  
    From the occupied molecular orbitals, the new density is constructed as
    \[
    D_{\mu\nu} = 2\sum_{a=1}^{N_{\text{occ}}} C_{\mu a} C_{\nu a}^{*},
    \]
    via \texttt{ComputeNewDensityMatrix()}.  
    The factor of two accounts for spin degeneracy in closed-shell systems.

    \item \textbf{Energy Evaluation and Convergence Test:}  
    The total electronic energy is evaluated as
    \[
    E_{\text{elec}} = \tfrac{1}{2}\,\mathrm{Tr}[\mathbf{D}(\mathbf{H}^{\text{core}} + \mathbf{F})].
    \]
    Convergence is achieved when
    \[
    \Delta E = |E^{(n)} - E^{(n-1)}| < \epsilon \quad \text{or} \quad 
    \|\mathbf{D}^{(n)} - \mathbf{D}^{(n-1)}\| < \epsilon,
    \]
    as checked in \texttt{CheckHFockEnergyConverg()}.
  \end{enumerate}

  \item \textbf{Final Output:}  
  Upon convergence, FSIM stores the results in the \texttt{ManagerResult} structure, providing access to the final electronic and total molecular energies.  The nuclear repulsion energy, obtained from the \texttt{Molecule} object, is added to yield the total molecular energy.
\end{enumerate}

The \texttt{HartreeFockEngine} uses three key inputs obtained from the molecular integral engine calculations: the core 
Hamiltonian $\mathbf{H}^{\text{core}}$, the overlap matrix $\mathbf{S}$, and the four-index electron-repulsion tensor $(\mu\nu|\lambda\sigma)$. These inputs are stored in Armadillo matrix and tensor structures for efficient linear algebra operations.

At the higher level, the \texttt{HartreeFockManager} class integrates the HF procedure into the FSIM computational pipeline.  
It sequentially:
\begin{enumerate}
  \item Parses the molecular input and basis set data via \texttt{CreateShellPairData()},
  \item Invokes the McMurchie--Davidson module to compute one- and two-electron integrals,
  \item Executes the SCF cycle through \texttt{RunHartreeFockManager()},
  \item Store the HF energy and relevant metadata.
\end{enumerate}

This modular architecture decouples integral generation from the SCF driver, allowing 
independent testing and profiling of both components.  
Despite its simplicity, the FSIM HF engine accurately reproduces reference results 
obtained with established quantum chemistry packages, as shown in Section \ref{sec:validation}.

\subsubsection{Direct Inversion in the Iterative Subspace (DIIS)}

To accelerate convergence of the SCF iterations, 
FSIM implements the \emph{Direct Inversion in the Iterative Subspace} (DIIS) method~\cite{PULAY1980393}. 
DIIS extrapolates a new Fock matrix as an optimal linear combination 
of previous iterates so as to minimize the commutator residuals of the HF equations.

At the $i$–th SCF iteration, we define the Fock matrix $\mathbf{F}_i$ in the atomic basis and the corresponding 
density matrix $\mathbf{D}_i$.  The commutator error (or DIIS residual) 
is given by the $\mathbf{e}_i$ matrices\footnote{The commutator error, which represents a deviation 
from the HF stationarity condition $[\mathbf{F}, \mathbf{DS}] = 0$. },
\begin{equation}
  \mathbf{e}_i = \mathbf{F}_i \mathbf{D}_i \mathbf{S} - \mathbf{S} \mathbf{D}_i \mathbf{F}_i,
  \label{eq:diis-error}
\end{equation}
where $S$ is the atomic–orbital overlap matrix.  The goal of DIIS is to obtain 
an extrapolated Fock matrix
\begin{equation}
  \mathbf{F}_{\mathrm{DIIS}} = \sum_{i=1}^{m} c_i \mathbf{F}_i,
  \label{eq:diis-fock}
\end{equation}
whose associated residual 
$\mathbf{e}_{\mathrm{DIIS}} = \sum_i c_i \mathbf{e}_i$ has the smallest possible norm,
subject to the constraint
\begin{equation}
  \sum_{i=1}^{m} c_i = 1.
  \label{eq:diis-constraint}
\end{equation}

Minimizing $\|e_{\mathrm{DIIS}}\|^2$ with the above constraint leads to the 
Lagrangian
\begin{equation}
  \mathcal{L} = 
  \sum_{i,j} c_i c_j \langle \mathbf{e}_i, \mathbf{e}_j \rangle
  - \lambda \left( \sum_i c_i - 1 \right),
\end{equation}
where 
$\langle \mathbf{e}_i, \mathbf{e}_j \rangle = \mathrm{Tr}(\mathbf{e}_i^\dagger \mathbf{e}_j)$
is the Frobenius inner product between residual matrices.
Setting the derivatives of $\mathcal{L}$ with respect to 
$c_i$ and the Lagrange multiplier $\lambda$ to zero yields the linear system
\begin{equation}
  \begin{pmatrix}
    \mathbf{B} & \mathbf{1} \\
    \mathbf{1}^T & 0
  \end{pmatrix}
  \begin{pmatrix}
    \mathbf{c} \\ \lambda
  \end{pmatrix}
  =
  \begin{pmatrix}
    \mathbf{0} \\ 1
  \end{pmatrix},
  \qquad
  \mathbf{B}_{ij} = \langle \mathbf{e}_i, \mathbf{e}_j \rangle.
  \label{eq:diis-system}
\end{equation}
Solving Eq.~\eqref{eq:diis-system} gives the coefficients $\{c_i\}$ defining 
the optimal extrapolated Fock matrix in Eq.~\eqref{eq:diis-fock}.

In the \texttt{DIISManager} class, FSIM constructs the overlap matrix $B$ 
from the commutator residuals of recent SCF iterations,
solves Eq.~\eqref{eq:diis-system} for $\mathbf{c}$,
and forms the new Fock matrix as
\begin{equation}
  F_{\mathrm{DIIS}} = \sum_i c_i F_i.
\end{equation}
The number of stored Fock/error pairs (typically 6–8) 
defines the dimension of the iterative subspace.
This procedure effectively ``inverts'' the iteration history to predict 
a Fock matrix that best cancels past residuals, yielding rapid and 
stable SCF convergence.

The following pseudocode summarizes the algorithm implemented 
in \texttt{DIISManager} for constructing the DIIS Fock matrix.  
The code corresponds directly to Eqs.~\eqref{eq:diis-error}–\eqref{eq:diis-system}.

\begin{algorithm}[H]
\caption{DIIS extrapolation of the Fock matrix}
\begin{algorithmic}[1]
  \STATE \textbf{Input:} Fock matrices $\{\mathbf{F}_i\}$ and error matrices $\{\mathbf{e}_i\}$, $i=1,\dots,m$
  \STATE \textbf{Output:} Extrapolated Fock matrix $F_{\mathrm{DIIS}}$
  \vspace{3pt}
  \STATE Construct matrix $\mathbf{B}$ of size $(m{+}1)\times(m{+}1)$:
  \FOR{$i,j = 1$ to $m$}
    \STATE $B_{ij} \gets \mathrm{Re}\!\left[ \mathrm{Tr}( \mathbf{e}_i^\dagger \mathbf{e}_j ) \right]$
  \ENDFOR
  \STATE Set $B_{i,m+1} = B_{m+1,i} = -1$ and $B_{m+1,m+1} = 0$
  \STATE Define right--hand side vector $\mathbf{b} = (0,0,\dots,0,-1)^T$
  \STATE Solve $\mathbf{B}\mathbf{x} = \mathbf{b}$ for $\mathbf{x} = (c_1,\dots,c_m,\lambda)^T$
  \STATE Form $\mathbf{F}_{\mathrm{DIIS}} = \sum_{i=1}^{m} c_i \mathbf{F}_i$
  \STATE \textbf{return} $F_{\mathrm{DIIS}}$
\end{algorithmic}
\end{algorithm}

In the FSIM implementation, the inner products 
$\mathrm{Tr}(e_i^\dagger e_j)$ are evaluated using Armadillo’s 
\texttt{cdot} operation on vectorized matrices, and 
the linear system in Eq.~\eqref{eq:diis-system} is solved 
with \texttt{arma::solve}.  
The extrapolated matrix $F_{\mathrm{DIIS}}$ replaces the current 
Fock matrix in the SCF cycle before diagonalization.

\begin{figure}[htbp]
  \centering
  \includegraphics[width=0.7\textwidth]{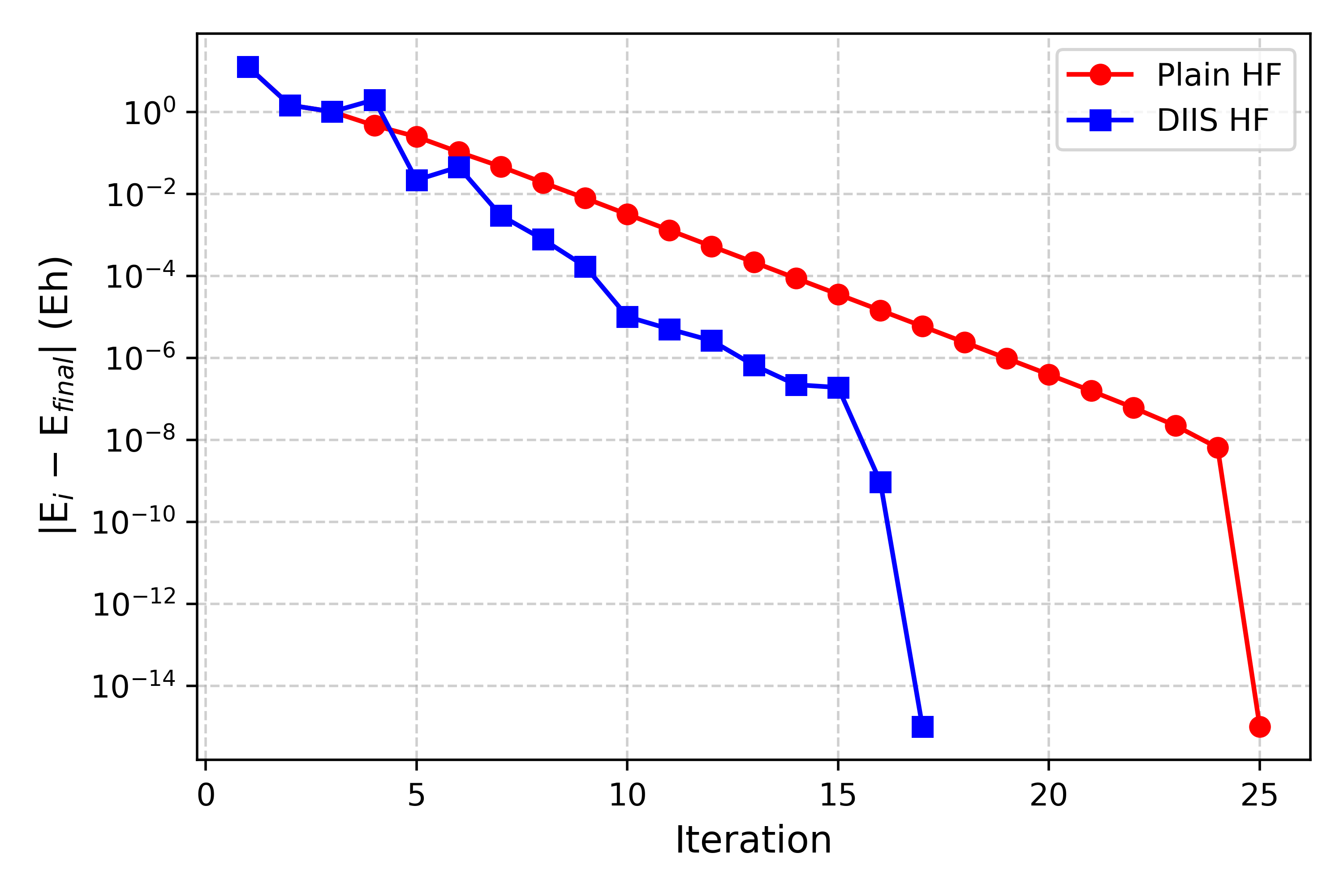}
  \caption{
    Comparison of SCF convergence for the HF method
    with and without DIIS acceleration for H2O and the aug-cc-pVDZ basis.
    The plot shows the logarithmic decay of the absolute energy difference
    $\left| E_i - E_\mathrm{final} \right|$ as a function of iteration number.  
  }
  \label{fig:scf-diis-convergence}
\end{figure}

The convergence behavior of the SCF procedure is shown in
Fig.~\ref{fig:scf-diis-convergence}. The plot compares the standard HF
iteration scheme and the DIIS approach for a water molecule slightly displaced 
from its equilibrium geometry. The vertical axis displays the logarithmic energy difference
$\left| E_i - E_\mathrm{final} \right|$, which measures how far each iteration $i$
is from the converged HF energy. The rapid drop of this quantity for
the DIIS method illustrates its ability to significantly improve convergence
stability and rate compared to conventional SCF.

\begin{figure}[!htbp]
  \centering
  \begin{subfigure}{0.7\textwidth}
    \centering
    \includegraphics[width=\linewidth]{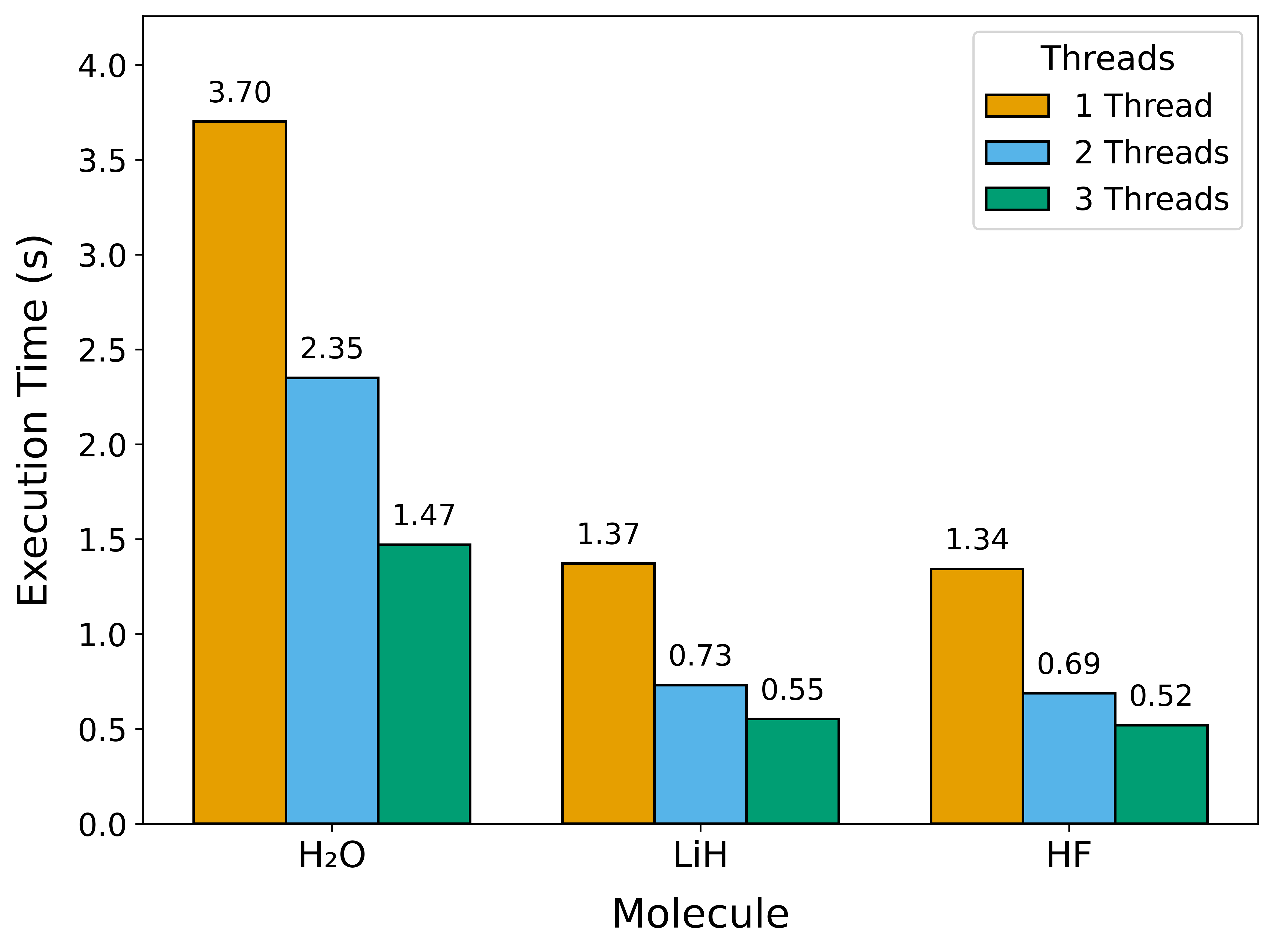}
  \end{subfigure}
  \vspace{1em} 
  \begin{subfigure}{0.7\textwidth}
    \centering
    \includegraphics[width=\linewidth]{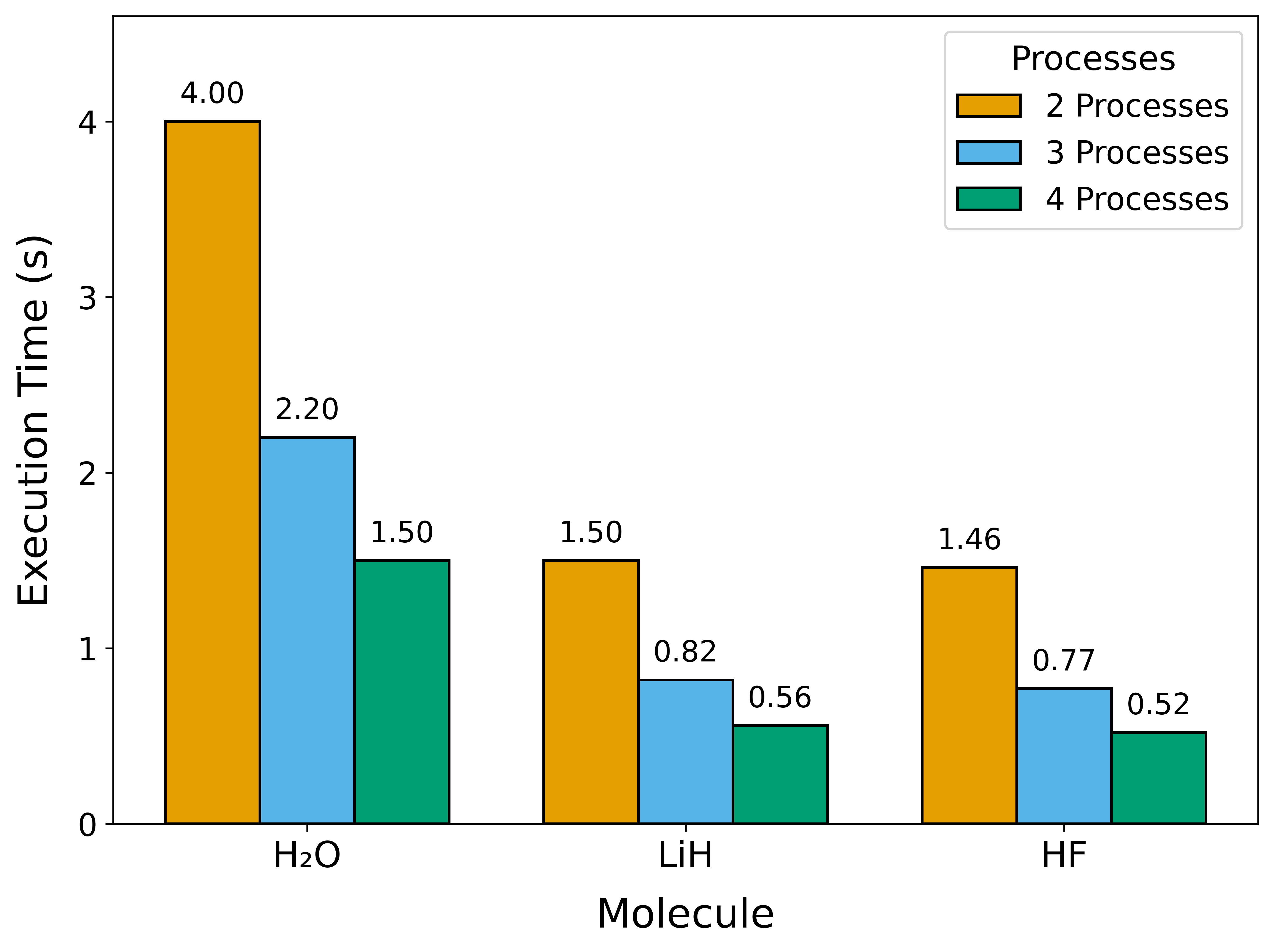}
  \end{subfigure}
  \caption{HF execution times for small molecular systems using different parallelization models. 
  The top panel shows results obtained with OpenMP using one to three threads, while the bottom panel shows 
  corresponding results with MPI using two to four processes. 
  Both plots illustrate how parallelization significantly reduces wall-clock time for small molecules and demonstrate the 
  impact of shared- versus distributed-memory parallelism within the FSIM framework.}
  \label{fig:scaling_comparison}
\end{figure}

\subsection{Integral Engine Implementation}

The computational core of FSIM relies on the McMurchie--Davidson formalism, which 
expands products of Gaussian functions into Hermite--Gaussian forms, allowing efficient 
recursive evaluation of all required one- and two-electron integrals. These include 
overlap, kinetic energy, nuclear attraction, and electron repulsion integrals, 
which are evaluated analytically using Hermite recurrence relations and Boys 
functions, as described in previous sections. 

In this work, we focus exclusively on the implementation of the electron 
repulsion integral (ERI) computation using MPI and OpenMP, while the implementations \
of other integral types are omitted to keep the paper concise.

\subsubsection{Parallel Evaluation of Integrals with MPI}

The computation of ERIs can be parallelized in FSIM using MPI to distribute the integral 
computation across several processes. Once the 
four-index tensor $(pq|rs)$ is assembled, it is stored  
and later used in the serial HF SCF procedure. 

The parallel computation, implemented in the class \texttt{HartreeFock\_MPI\_Manager}, adopts a 
manager–-worker model as described in Ref.~\cite{Janssen2008ParallelComputing}. 
All MPI processes execute the same code, but  their roles differ depending on their rank:
\begin{itemize}
  \item \textbf{Manager (rank 0)} creates the list of shell-pair tasks, distributes them dynamically 
  to the workers, and gathers the final results.
  \item \textbf{Worker processes (ranks $>0$)} receive shell indices $(m,n)$, compute the corresponding 
  set of unique two-electron integrals $(m,n,r,s)$ using the \texttt{MolecularIntegralEngine}, and contribute 
  their partial results to the construction of the tensor $g_{pqrs}$.
\end{itemize}

Each task corresponds to a unique set of integrals $(m,n|r,s)$, where the shell pair $(m,n|$ is fixed, and 
the remaining shell pairs $|r,s)$ are determined at runtime by the worker within the computing engine layer 
using permutation symmetries.
The manager builds this task list and sends work to idle workers on demand.  
This approach ensures dynamic load balancing, as the computational cost can vary significantly 
among shell pairs. Communication between processes is managed through three MPI message tags.
The tag \texttt{TASK\_REQUEST} is sent by a worker process to signal that it is ready to receive 
a new task. In response, the manager sends a message with the tag \texttt{TASK\_SEND}, which 
contains the corresponding pair of shell indices $(m,n)$. Finally, when no tasks remain, the manager
broadcasts a \texttt{TERMINATE} message to indicate that all processes should stop execution.

\begin{lstlisting}[
  language=C++,
  basicstyle=\ttfamily\scriptsize,
  frame=single,
  caption={Manager routine for dynamic task distribution.},
  label={lst:dynamic-routine}
]
while (num_workers > 0) {
    MPI_Recv(&worker_rank, 1, MPI_INT, MPI_ANY_SOURCE,
             TASK_REQUEST, MPI_COMM_WORLD, &status);
    if (task_index < num_tasks) {
        MPI_Send(tasks[task_index].data(), 2, MPI_INT,
                 worker_rank, TASK_SEND, MPI_COMM_WORLD);
        task_index++;
    } else {
        MPI_Send(NULL, 0, MPI_INT, worker_rank,
                 TERMINATE, MPI_COMM_WORLD);
        num_workers--;
    }
}
\end{lstlisting}

\begin{lstlisting}[
  language=C++,
  basicstyle=\ttfamily\scriptsize,
  frame=single,
  caption={Worker routine for distributed ERI computation.},
  label={lst:worker-routine}
]
while (true) {
    MPI_Send(&rank, 1, MPI_INT, 0, TASK_REQUEST, MPI_COMM_WORLD);
    MPI_Recv(task, 2, MPI_INT, 0, MPI_ANY_TAG, MPI_COMM_WORLD, &status);
    if (status.MPI_TAG == TERMINATE) break;
    md_algo.DriveCompPrimitiveEris(task[0], task[1]);
}
\end{lstlisting}

The code fragments in Listing~\ref{lst:dynamic-routine} and \ref{lst:worker-routine} illustrate 
the core communication pattern of the manager and worker routines.

Each worker computes its assigned ERIs using the McD 
recursion scheme, which efficiently generates all required primitive integrals for a given shell pair. 
Once all tasks are completed, the manager process collects the distributed data through a global 
reduction operation that sums the contributions from the partially filled four-index ERI tensors:
\[
\mathbf{g} = \sum_{r=0}^{N_{\text{ranks}}-1} \mathbf{g}_r^{(\text{partial})},
\]
implemented in the code via the \texttt{MPI\_Reduce} routine. 
The resulting tensor, containing the complete set of molecular integrals, is then written to disk 
and subsequently used in the HF self-consistent field procedure.

This implementation provides a simple but effective approach for distributing the most expensive part of the computation across multiple nodes.  
Dynamic task allocation ensures balanced workload, while communication overhead remains minimal because only small messages (shell indices and integral buffers) are exchanged.  
Once the complete $g_{pqrs}$ tensor is assembled, the SCF iterations proceed serially using the precomputed integrals.

\subsubsection{Parallel Evaluation of Integrals with OpenMP}

For shared-memory parallelism, the evaluation of the two-electron repulsion integrals (ERIs) is 
parallelized using the OpenMP API.  
In contrast to the MPI version, which distributes shell pairs among processes, the OpenMP implementation performs 
parallel work sharing among threads within a single node. This strategy efficiently utilizes all available 
CPU cores without the need for inter-process communication.

\begin{lstlisting}[language=C++, basicstyle=\ttfamily\scriptsize, frame=single,
caption={Parallel evaluation of electron-repulsion integrals (ERIs) using OpenMP.
Each thread processes a distinct subset of canonical shell quartets, computing the associated Hermite Coulomb integrals in parallel.
The resulting partial tensors are accumulated into the shared t\_primitive\_ERIs array.}, 
label={lst:openmp_implementation}]
void MolecularIntegralEngine::ComputePrimitiveERIs() 
{
  const auto& shell_pair_matrix = shell_pair_data_.shell_pair_matrix;
  const int num_shell = shell_pair_data_.get_num_shells() - 1;
  CanonicalIndices<int> canon_ids_container;
  ComputeCanonicalTuples(canon_ids_container, 0, num_shell);
  Sort4Shells(canon_ids_container.canonical_ids);

  #pragma omp parallel shared(canon_ids_container, shell_pair_matrix, t_contracted_eris)
  {
    #pragma omp for schedule(monotonic : dynamic, 1) nowait
    for (const auto& canonical_indices: canon_ids_container.canonical_ids){
      auto [m, n, r, s] = canonical_indices; 
      const ShellPair& sp_mn = shell_pair_matrix(m, n);
      const ShellPair& sp_rs = shell_pair_matrix(r, s);
      HermiteCoulombIntegralCalculator hc_int(sp_mn, sp_rs);
      hc_int.ReserveMemory();
      hc_int.ComputeTensors();
      hc_int.ComputeHmiteClombIntegrals();
      
      TwoElectronIntsCalculatorCompact two_elect_ints(hc_int.tHermite, hc_int.tHermite_compact, hc_int.t_alpha, sp_mn, sp_rs, t_contracted_eris);
      two_elect_ints.ComputeIntegralSet();
    }
  }
}
\end{lstlisting}

The OpenMP parallelization is implemented within the \texttt{MolecularIntegralEngine} class, specifically 
in the \texttt{ComputePrimitiveERIs()} routine. In this routine, the integrals are evaluated 
over \emph{canonical} shell quartets $(m,n,r,s)$, which represent unique combinations of 
atomic orbital shells. Before entering the parallel region, a list of canonical index tuples 
is generated to exploit permutation symmetries and is sorted in ascending order of estimated 
computational cost to improve load balancing as much as possible.
The core structure of the parallel routine is illustrated in Listing~\ref{lst:openmp_implementation}.

The \texttt{\#pragma omp parallel} directive creates a team of threads that share access to 
the integral data structures.  
Each thread independently processes a subset of the canonical shell quartets, invoking 
the McMurchie--Davidson recursion relations to compute all corresponding Hermite integrals 
and primitive ERIs. The shared tensor \texttt{t\_contracted\_eris} accumulates the 
computed integrals in memory.

The loop is scheduled dynamically using
\begin{verbatim}
#pragma omp for schedule(monotonic : dynamic, 1)
\end{verbatim}
which assigns one shell quartet at a time to each thread.  
This scheduling strategy minimizes idle time and improves load balancing, as the cost of 
integral evaluation varies significantly with the angular momentum and contraction length 
of the shells involved. For each assigned shell quartet, every thread retrieves the corresponding 
shell pairs $(m,n)$ and $(r,s)$ from the shell-pair matrix and constructs the necessary Hermite 
tensors and intermediate quantities using the \texttt{HermiteCoulombIntegralCalculator}. The Hermite 
Coulomb integrals are then evaluated and accumulated in an intermediate tensor and the 
full $(m n | r s)$ integral block is computed via \texttt{ComputeIntegralSet}. Since all threads 
operate within a shared-memory environment and store ERI values in separate memory regions, 
synchronization is minimal. The use of the \texttt{nowait} clause further reduces synchronization 
overhead by allowing threads to proceed without waiting at the end of the parallel loop.

Fig.~\ref{fig:scaling_comparison} show the execution times of the full HF workflow 
using the OpenMP and MPI parallel implementations, respectively. 
For the small molecular systems tested, both approaches demonstrate clear reductions in wall-clock time as the number 
of threads or processes increases—typically by factors of two or more relative to the single-threaded case. 
These results illustrate the effectiveness of shared- and distributed-memory parallelization even at modest system 
sizes, while also highlighting the practical limits of strong scaling for small workloads. 
At present, MPI parallelization in \textsc{FSIM} is applied only to the molecular-integral evaluation stage, 
whereas the OpenMP implementation also accelerates the SCF procedure. 
This asymmetry explains the somewhat higher efficiency observed in the OpenMP runs. 
Beyond performance validation, these experiments serve as pedagogical demonstrations 
of how computational workload distribution, communication overhead, and algorithmic structure jointly 
determine parallel efficiency in electronic-structure codes.

\subsection{Testing and Validation}
\label{sec:validation}

A comprehensive testing and validation framework was developed to ensure both the 
numerical correctness and performance of the FSIM library, leveraging the CMake 
testing infrastructure (CTest) for automated build and test management.
The system combines integration tests for algorithmic verification with benchmark 
tests for performance.  
All computed HF energies and integral results are validated against 
reference data from established quantum chemistry packages such as Q-Chem \cite{Epifanovsky2021} and 
PySCF \cite{Sun2018}.
    
The testing infrastructure is organized into three main components:
\begin{itemize}
  \item \textbf{Integration Tests:} Validate correctness of parallel and serial HF and integral computations by comparing FSIM results with reference energies. 
  \item \textbf{Benchmark Tests:} Measure execution times of parallel and serial workflows, collect performance data, and store results in timestamped benchmark files.  
\end{itemize}

Integration tests employ strict \texttt{assert()} checks to compare FSIM results against 
reference HF molecular energies for a range of molecules and basis sets. 
Typical test systems include Be, H$_2$, LiH, HF, and H$_2$O, with basis sets spanning 
from minimal (STO-3G) to extended (aug-cc-pVDZ). In particular, parallel MPI tests 
validate the distributed integral computation routines. When MPI tests are triggered, only 
the root process (rank 0) performs validation and records timing data. When comparing 
with established software, numerical tolerances for both parallel and serial 
integration tests typically range from $10^{-6}$ to $10^{-8}$.

Benchmark tests, implemented in \texttt{benchmark\_test.cpp}, measure the wall-clock execution 
time of each major algorithmic component.  
The \texttt{BenchmarkDriver} class manages test execution, collects timing data using the \texttt{Chronometer} utility, and writes structured results to files of the form:
\[
\texttt{benchmark\_results\_[YYYY-MM-DD\_HH-MM-SS].txt.}
\]
Each benchmark entry stores execution time, molecule, basis set, and reference comparison 
data through the \texttt{HFBenchmarkEntry} and \texttt{BenchTestResult} data structures. 
Benchmark data are stored in a structured directory under \texttt{test/benchmark\_data/}, allowing 
easy post-processing and performance tracking over multiple runs.  
Results can be accumulated over time for regression analysis and scalability studies. 

To ensure reproducibility and automate validation, all tests are integrated into a continuous integration (CI) 
workflow defined in the project’s \texttt{.gitlab-ci.yml} file. The CI pipeline 
runs within a Docker container configured with the necessary MPI and build tools, and consists 
of two stages: \texttt{build} and \texttt{test}. The build stage compiles the 
library using CMake (with optional MPI support), while the test stage executes all registered tests 
through \texttt{ctest -V}. Artifacts from the build stage are passed automatically 
to the testing stage, enabling fully automated verification of each commit and merge request. 
This CI setup guarantees that FSIM remains numerically consistent and build-stable across code revisions.

As an example of the agreement between results obtained with different packages, Table~\ref{tab:energies_accuracy} summarizes 
the HF total energies computed with \textsc{FSIM} for several representative molecules and compares them 
with reference values from \textsc{PySCF} and \textsc{Q-Chem}.
The excellent agreement—matching to within numerical precision for different basis sets—confirms 
the correctness of the integral evaluation and self-consistent field procedures implemented in \textsc{FSIM}. 
Beyond numerical validation, these benchmarks serve a pedagogical role by illustrating how independent 
implementations based on the same theoretical foundations can be cross-checked for accuracy.

\begin{table}[t]
\centering
\caption{
Comparison of HF total energies (in Hartree) computed 
using \textsc{FSIM}, \textsc{PySCF}$^{(a)}$, and \textsc{Q-Chem}$^{(b)}$ 
for selected molecules with STO-3G and aug-cc-pVDZ basis sets.  The HF 
self-consistent field convergence threshold was set to $10^{-6}$ Hartree.
}
\label{tab:energies_accuracy}
\begin{tabular}{lrcc}
\hline
Molecule & Basis &  \textbf{$E_{\text{FSIM}}$} & \textbf{$E_{\text{External}}$} \\
\hline
LiH       & STO-3G        & -7.8620020     & -7.8620020$^{a}$      \\
          & aug-cc-pVDZ   & -7.9841442     & -7.9841442$^{b}$       \\
HF        & STO-3G        & -98.570757     & -98.570757$^{b}$       \\
          & aug-cc-pVDZ   & -100.033474     & -100.033474$^{b}$       \\
H$_2$O    & STO-3G        & -74.962991     & -74.962991$^{b}$       \\
          & aug-cc-pVDZ   & -76.0414047     & -76.0414047$^{b}$       \\
\hline
\end{tabular}
\begin{flushleft}
\footnotesize
$^{(a)}$ Computed using \textsc{Q-Chem}. \\
$^{(b)}$ Computed using \textsc{PySCF}.
\end{flushleft}
\end{table}

\subsection{Performance Analysis and Profiling}
\label{sec:profiling}

Performance profiling is a fundamental aspect of high-performance software development and should 
form an integral, accessible component of any scientific code. In \textsc{FSIM}, profiling is treated 
not only as a tool for optimization but also as a learning aid that helps users understand where computational 
effort is spent and how parallel strategies affect performance. Currently, the framework includes simple, automated 
profiling scripts based on the GNU tools \texttt{gprof} and \texttt{gprofng}, which assist in identifying
computational bottlenecks and evaluating parallelization strategies in both the HF and 
molecular–integral evaluation routines. These tools allow users to measure sequential and 
parallel performance (under OpenMP), quantify computational loads across critical modules, and analyze 
memory usage patterns—providing immediate feedback for experimentation and tuning.

\begin{figure}[tbp]
\centering
\begin{minipage}[t]{0.47\textwidth}
\centering
\captionof{table}{CPU utilization from \texttt{gprofng} profiling. 
Times are exclusive CPU times normalized to the total runtime of 18.643~s.}
\label{tab:gprofng_cpus}
\begin{tabular}{lcc}
\hline
\textbf{CPU} & \textbf{Time (s)} & \textbf{(\%)} \\
\hline
5 & 6.705 & 35.96 \\
6 & 4.023 & 21.58 \\
0 & 3.763 & 20.18 \\
3 & 3.422 & 18.36 \\
2 & 0.410 & 2.20 \\
4 & 0.290 & 1.56 \\
1 & 0.030 & 0.16 \\
\hline
\textbf{Total} & 18.643 & 100.00 \\
\hline
\end{tabular}
\end{minipage}
\hfill
\begin{minipage}[t]{0.47\textwidth}
\centering
\captionof{table}{Thread-level distribution of exclusive CPU time from \texttt{gprofng}. 
Most workload is concentrated in threads 4–5, corresponding to intensive OpenMP regions.}
\label{tab:gprofng_threads}
\begin{tabular}{lcc}
\hline
\textbf{Thread} & \textbf{Time (s)} & \textbf{(\%)} \\
\hline
5 & 7.115 & 38.16 \\
4 & 6.645 & 35.64 \\
3 & 3.763 & 20.18 \\
1 & 0.610 & 3.27 \\
2 & 0.510 & 2.74 \\
\hline
\textbf{Total} & 18.643 & 100.00 \\
\hline
\end{tabular}
\end{minipage}

\caption{Summary of CPU and thread utilization obtained from \texttt{gprofng} profiling of a single point 
HF energy calculation for LiH with the aug-cc-pVTZ basis. These results highlight the imbalance typical of OpenMP parallel regions 
and provide a pedagogical example of how profiling can guide load-balancing optimizations.}
\label{fig:gprofng_summary}
\end{figure}

Profiling with \texttt{gprof} results show that the contractions used for computing ERIs [see Eq.~\ref{eq:eri-tensor-contracted}] 
and the construction of intermediate Hermite integral tensors [Eq.~\ref{eq:Rn-definition}] dominate the total 
runtime for small molecules. These components therefore serve as clear pedagogical examples and 
practical targets for performance optimization. 

Profiling with \texttt{gprofng} also provides a detailed breakdown of CPU and thread 
utilization during the HF workflow (Tables~\ref{tab:gprofng_cpus} 
and~\ref{tab:gprofng_threads}). The results reveal an uneven distribution of computational 
load across the available processing units, reflecting the intrinsic challenges of evaluating 
shell quartets with varying angular momenta and contraction coefficients.
These profiling measurements serve as a clear pedagogical example of how performance-analysis 
tools can be used to identify load imbalance and guide the development of more efficient parallel 
strategies.

By integrating profiling into both the code and the learning process, \textsc{FSIM} ensures that algorithmic 
design decisions are informed by quantitative data. 
This approach supports not only sustained optimization but also the development of a deeper understanding 
of performance engineering in computational chemistry.

\section{Pedagogical Extensions and Mini-Projects}
\label{sec:projects}

To encourage active learning, we propose a set of open-ended 
mini-projects based on the current FSIM implementation. These activities are designed 
to deepen understanding of high-performance computing concepts while contributing to the 
continued development of the codebase. They may serve as starting points for student 
projects, training exercises, or exploratory research in computational chemistry and 
scientific software engineering.

\begin{itemize}
    \item \textbf{Project 1: Hybrid Parallelization.} 
    Extend the existing MPI and OpenMP infrastructure to support hybrid execution within 
    distributed nodes, analyzing performance and scalability across different system architectures.

    \item \textbf{Project 2: Asynchronous Communication.} 
    Implement non-blocking MPI operations to overlap computation and communication, 
    and evaluate the resulting performance gains on representative molecular systems.

    \item \textbf{Project 3: Memory and Data Layout Optimization.} 
    Redesign key data structures to improve cache locality and memory throughput, 
    comparing results with the baseline FSIM implementation.

    \item \textbf{Project 4: GPU Acceleration.} 
    Integrate a GPU-based kernel using CUDA, HIP, or SYCL to offload computationally 
    intensive integral evaluations, and benchmark its efficiency relative to CPU-only runs.

    \item \textbf{Project 5: Benchmarking and Validation Suite.} 
    Develop a reproducible, domain-specific benchmarking framework for automated testing 
    of accuracy, performance, and scalability across different hardware configurations.
\end{itemize}

Each project emphasizes both theoretical understanding and practical implementation, offering opportunities for hands-on experience in algorithmic optimization, parallel programming, and performance analysis. 
By pursuing these extensions, contributors can help evolve FSIM into a richer platform for research and education in computational chemistry and high-performance computing.

\section{Future Directions and Learning Opportunities}

This work has presented a minimal yet extensible C++ implementation of the 
HF method and the McD scheme for molecular integral evaluation, 
developed within HPC-oriented framework. 
Throughout the paper, we have emphasized both algorithmic transparency and pedagogical clarity, 
illustrating how theoretical principles map directly onto efficient computational design. 
These foundations establish a platform for continued exploration, teaching, and development of the FSIM library.

Achieving greater scalability and efficiency within the FSIM framework offers an opportunity to explore advanced 
strategies used in high-performance codes while deepening understanding of algorithmic design principles. 
Future developments may include implementing \emph{direct self-consistent field} algorithms 
\cite{1982Almlof, Truhlar2000}, in which molecular integrals are recomputed on demand during each Fock 
matrix update to reduce memory usage and improve performance for large systems. 
Incorporating integral \emph{screening} techniques and \emph{density fitting} approximations \cite{2012Reine}—standard 
approaches in modern quantum-chemistry software—would further decrease the computational cost of two-electron repulsion 
integrals. These enhancements also provide rich pedagogical opportunities: contributors can refine data 
structures, experiment with template metaprogramming to accelerate integral evaluation, and optimize tensor 
operations for parallel architectures, thereby gaining hands-on experience with both algorithmic and 
architectural aspects of high-performance scientific computing.
With these extensions, FSIM will continue to evolve into a comprehensive high-performance and educational 
platform for computational chemistry.

\bibliographystyle{unsrt}
\bibliography{references}

\end{document}